\documentclass[twocolumn,preprintnumbers,prmaterials]{revtex4-2}

\usepackage[utf8]{inputenc}
\usepackage[colorlinks=true,allcolors=blue!80!black]{hyperref}
\usepackage{amsmath}
\usepackage[caption = false]{subfig}
\usepackage{graphicx,epstopdf}
\usepackage{blindtext}
\usepackage{lipsum}
\usepackage{amsfonts}
\usepackage{bbm}
\usepackage{amssymb}
\usepackage{enumerate}
\usepackage{color}
\usepackage{latexsym}
\usepackage{physics}
\usepackage{times,txfonts}
\usepackage{xcolor}
\usepackage{epsfig}
\usepackage{bm}
\usepackage{tabularx}
\usepackage{multirow}
\usepackage{mathtools}
\usepackage{xfrac}
\usepackage{physics}
\usepackage{enumerate}
\usepackage{braket}
\usepackage{soul,color}
\soulregister\cite7
\soulregister\ref7
\soulregister\pageref7
\setstcolor{red}
\usepackage{booktabs}
\usepackage{longtable}
\newcolumntype{F}[1]{%
    >{\raggedright\arraybackslash\hspace{0pt}}p{#1}}%
\newcolumntype{C}[1]{%
    >{\centering\arraybackslash\hspace{0pt}}p{#1}}%
\newcommand{\gw}{$G_0W_0$@PBEsol}
\newcommand{\gv}{$G_0W_0$}

\begin{document}

\title{A benchmark of first-principles methods for accurate prediction of semiconductor band gaps}

\author{Saeid Abedi}
\email{saeid.abedi@ph.iut.ac.ir}
\affiliation{Department of Physics, Isfahan University of Technology, Isfahan, 84156-83111, Iran}
\author{Mehdi Tarighi Ahmadpour}
\affiliation{Department of Physics, Isfahan University of Technology, Isfahan, 84156-83111, Iran}
\author{Samira Baninajarian}
\affiliation{Department of Physics, Isfahan University of Technology, Isfahan, 84156-83111, Iran}
\author{Hamideh Kahnouji}
\affiliation{Department of Physics, Isfahan University of Technology, Isfahan, 84156-83111, Iran}
\author{S. Javad Hashemifar}
\email{hashemifar@iut.ac.ir}
\affiliation{Department of Physics, Isfahan University of Technology, Isfahan, 84156-83111, Iran}
\author{Zhong-Kang Han}
\affiliation{Skolkovo Institute of Science and Technology, Bolshoy Boulevard 30/1, 121205 Moscow, Russia}
\author{Sergey V. Levchenko}
\email{s.levchenko@skoltech.ru}
\affiliation{Skolkovo Institute of Science and Technology, Bolshoy Boulevard 30/1, 121205 Moscow, Russia}

\date{\today} 

\begin{abstract}
  The band gap is an important parameter of semiconductor materials that influences several functional properties, in particular optical properties. However, a fast and reliable first-principles prediction of band gaps remains a challenging problem. Standard DFT approximations tend to strongly underestimate band gaps, while the more accurate $GW$ and hybrid functionals are much more computationally demanding and unsuitable for high-throughput screening. In this work, we have performed an extensive benchmark of several approximations with different computational complexity (\gw, HSE06, PBEsol, mBJ, PBEsol$-1/2$, and ACBN0) to evaluate and compare their performance in predicting the band gap of semiconductors. The benchmark is based on 114 binary semiconductors of different compositions and crystal structures, where about half of them have experimental band gaps. We find that, as expected, \gw\ performs well relative to the experiment, with a noticeable underestimation of the band gaps by about 14\%\ on average. Surprisingly, \gw\ is followed closely by the much computationally cheaper pseudo-hybrid ACBN0 functional, showing an excellent performance with respect to experimental data. The meta-GGA mBJ functional also performs well relative to the experiment, even slightly better than \gw\ in terms of mean absolute (percentage) error. The HSE06 and PBEsol$-1/2$ schemes perform overall worse than ACBN0 and mBJ schemes but much better than PBEsol. Comparing the calculated band gaps on the whole data set (including the samples with no experimental band gap), we find that HSE06 and mBJ have excellent agreement with respect to the reference \gw\ band gaps. Thus, we propose the mBJ band gaps as economic descriptors when developing artificial intelligence models for high-throughput screening of semiconductor band gaps.
\end{abstract}

\maketitle

\section*{Introduction}
A reliable prediction of the fundamental band gap of semiconductors and insulators, in contrast to their ground-state properties, is a challenging task for the Kohn-Sham density-functional theory (DFT) \cite{Kohn1965Self,Hohenberg1964Inhomogeneous}. It has been demonstrated recently that, although exact Kohn-Sham DFT does not reproduce fundamental gap, approximate DFT can in fact do it \cite{John2017Understanding}. However, the accuracy of calculated band gaps is sensitive to the choice of exchange-correlation (xc) functional approximation. Conventional approximations, including the local density (LDA) \cite{Perdew1981Self} and generalized-gradient approximations (GGA) of Perdew, Burke, and Ernzerhof (PBE) \cite{Perdew1996Generalized}, are not able to provide accurate band gaps because of the self-interaction error \cite{Perdew1981Self} and the absence of the derivative discontinuity \cite{Perdew1983Physical,Sham1983Density,Perdew1982Density}. The hybrid Heyd-Scuseria-Ernzerhof (HSE) functional \cite{Heyd2003Hybrid,Heyd2006Erratum}, meta-generalized gradient approximations (meta-GGAs) \cite{Sun2015Strongly,sun2016accurate,Sala2016Kinetic}, DFT+\textit{U} \cite{Liechtenstein1995Density,Anisimov1997First}, DFT$-1/2$ \cite{Ferreira2008Approximation, Ferreira2011Slater,XUE2018Improved,Doumont2019Limitations}, dynamical mean-field theory \cite{Georges1992Hubbard,Georges1996Dynamical,Kotliar2006Electronic} and the many-body Green's function based $GW$ methods \cite{Aryasetiawan1998The,Fuchs2007Quasiparticle,Shishkin2007Self} are developed to address this problem.

A benchmark study on a group of elemental and binary semiconductors indicated a maximum error of about 0.7\,eV in the band gaps obtained with the hybrid HSE03 functional, which is significantly better than the maximum error of more than 2\,eV (approximately 50\% for the relative error) for the conventional PBE functional \cite{Heyd2005Energy}. However, the non-local exact exchange term in the hybrid DFT schemes significantly increases the computational cost. Furthermore, the contribution of the exact exchange is controlled by an empirical hyper-parameter in this approach. Our results give a mean absolute error of about 0.8\,eV within HSE06. A more sophisticated scheme is the $GW$ method, which is believed to be the most accurate approach for calculating the quasiparticle band gap of semiconductors and insulators \cite{Aryasetiawan1998The}. However, this method is computationally very demanding \cite{Rinke2005Combining} and thus is not suitable for high-throughput studies. In the current work, a mean absolute error of about 0.6\,eV is predicted for this sophisticated scheme.

On the other hand, the DFT+$U$ approach provides a faster alternative for reliable prediction of the semiconductor band gaps. In this method, proper Hubbard $U$ parameters provide the opportunity to correct the calculated band gaps relative to the experiment or more expensive methods, with a low computational cost comparable to the conventional DFT calculations. Nevertheless, the determination of the proper Hubbard parameters requires a nontrivial effort in this approach, involving additional DFT calculations in the framework of random-phase approximation (RPA) \cite{Springer1998Frequency} and linear response approaches \cite{Cococcioni2005Linear}. Recently, an alternative method was proposed for on-the-fly self-consistent computation of the Hubbard parameters -- the ACBN0 pseudohybrid density functional. In ACBN0, a renormalization of the density matrix is employed to express the Hubbard parameters in terms of electron density to be determined during the self-consistent DFT calculations. We found no benchmark study of the band gaps within the DFT+$U$ scheme in the literature, which is likely because of the elaborating task of $U$ determination.

The modified Becke-Johnson potential (mBJ)\cite{Tran2006Accurate} is another fast method for calculating the band gap of bulk solids. This method has been shown to have mean absolute errors between {0.47-0.5}\,eV ({15-30\%} for the absolute relative error) when compared with experimental measurements \cite{Borlido2019Large,Tran2019Semilocal}. Our data indicate a high consistency between the mBJ and $G_{0}W_{0}$ methods with slightly better performance of mBJ in reproducing the experimental data.

Another computationally cheap approach for reliable band gap prediction is DFT$-1/2$ \cite{Ferreira2008Approximation}, which is a generalization of Slater's half-occupation (transition state) technique \cite{Slater1972Statistical, Slater1972Self} for periodic solids by means of adding a local potential to the Kohn-Sham (KS) potential. This method provides approximate quasiparticle corrections to the band structure and consequently improves the band gap for a wide range of semiconductors and insulators at a low computational cost comparable to the standard DFT calculations \cite{Ferreira2008Approximation}. The method has been applied to a number of test sets \cite{Ferreira2011Slater,Ferreira2013The,Ronaldo2017The} and has been shown to have a mean absolute error (MAE) of 0.44\,eV for LDA$-1/2$ and 0.67\,eV for PBE$-1/2$ \cite{Doumont2019Limitations}. It also has been shown as a good starting point for $G_{0}W_{0}$ calculations \cite{Rodrigues2016Probing}.

\begin{figure}[t]
    \centering
    \includegraphics[scale=0.25]{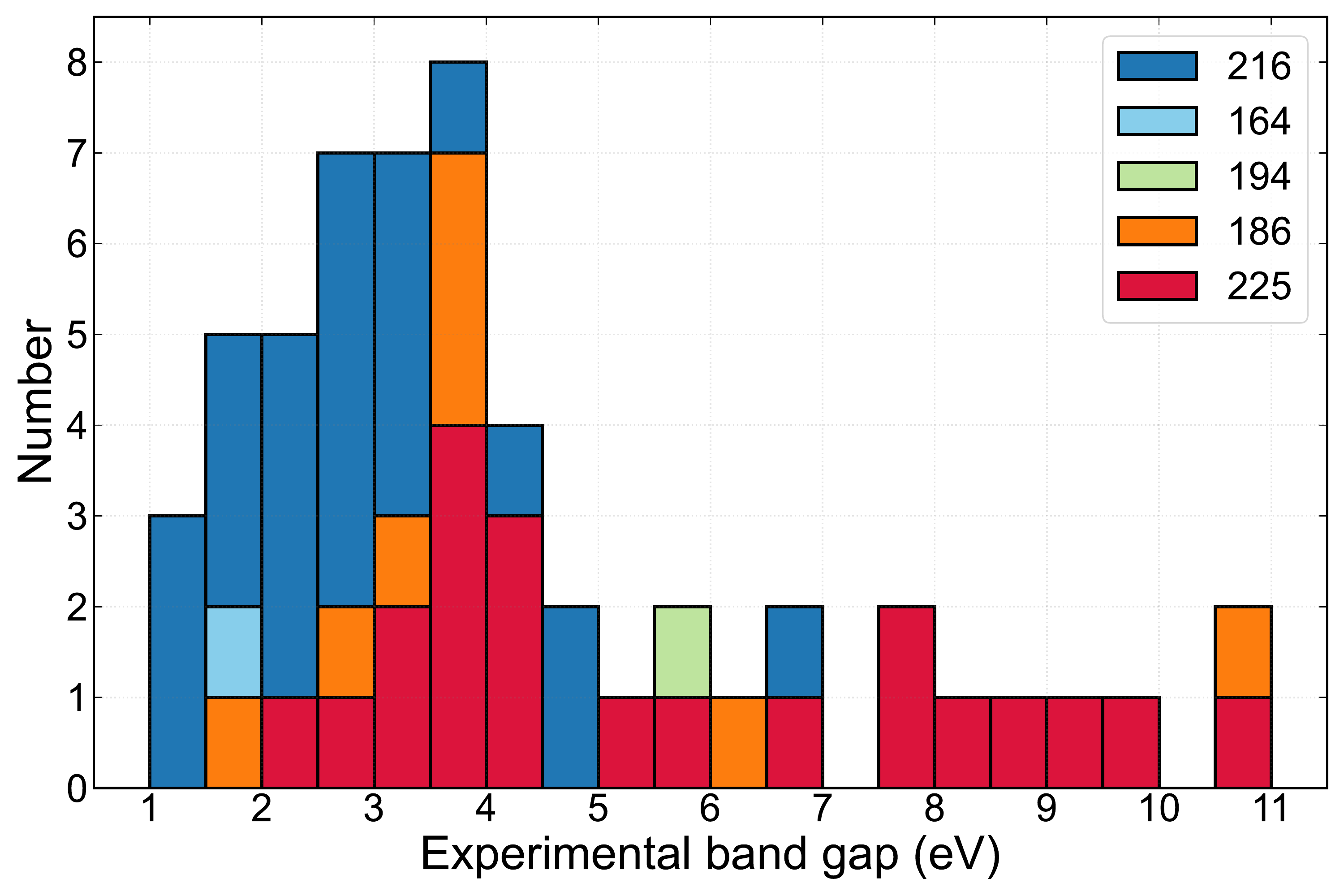}
    \caption{
             Experimental band-gap distribution histogram of investigated binary compounds. 
	     Different colors represent different crystal structures. 
	     The space group numbers 216, 164, 194, 186, and 225 indicate
	     F{$\bar4$}3m (Zincblende), P$\bar3$m1 (Trigonal), P6$_3$/mmc (Hexagonal), P6$_3$mc (Wurtzite), Fm3m (Rocksalt) structures. 
            } 
 \label{fig:histogram_exp} 
\end{figure}

In this paper, we report a benchmark study of band gaps of non-metallic compounds (with different structures) to compare the performance of a set of xc functionals, including ACBN0, HSE06, \gw, mBJ, and PBEsol$-1/2$ methods. In order to construct our first-principles benchmark data set, we randomly selected a group of binary semiconductors from the Materials Project database \cite{de2015charting}. The final data set has a size of 114 samples, where 56 compounds have experimental band gaps. The distribution of experimental band gaps is depicted in Fig.~\ref{fig:histogram_exp}. It can be seen that the majority of the gaps lie between 2 and 5\,eV, and some materials have a value of around 6\,eV to more than 10\,eV. First, we focus on the samples with experimental band gaps to evaluate the selected methods with reference to the experimental data. Then, the evaluation is extended to the whole data set by adapting the calculated \gw\ and ACBN0 band gaps as reference data.

\section*{Methods}
 For each material in the data set, geometry optimizations were performed in the framework of KS DFT by using the all-electron full-potential numeric atom-centered orbital approach \cite{Blum2009Ab} implemented in the Fritz Haber Institute ab-initio materials simulations (FHI-aims) program package. Numerical tabulation of atomic orbitals significantly enhances the flexibility of the basis functions and thus remarkably improves the efficiency of the DFT calculations with this package. We used the GGA PBEsol functional \cite{Perdew2008Restoring} for obtaining the relaxed geometries. This functional is a revised parameterization of PBE for a more accurate calculation of surface and crystal properties and especially the lattice constants of solids. The basis set and numerical grids were defined by ``tight'' default integration grid settings. We employed convergence criteria of $10^{-4}$\,eV, $10^{-7}$\,eV, and $10^{-5}$ {eV/\AA} for the sum of KS eigenvalues, the total energy, and the atomic forces, respectively. For sampling of the Brillouin zone, $\Gamma$-centered Monkhorst-Pack {\bf k}-point grids were used such that the smallest distance between grid points in reciprocal space became less than or equal 0.15. This {\bf k}-point sampling method resulted in a variable {\bf k} mesh for materials with different structures and lattice vectors. Localized basis functions in the FHI-aims package allow for a very efficient implementation of the hybrid functionals for crystalline systems \cite{Levchenko2015Hybrid}. Therefore, we used the FHI-aims package to calculate the electronic structure with the HSE06 functional. The HSE06 calculations were conducted with the tight setting and convergence criteria of $10^{-3}$\,eV and $10^{-6}$\,eV for the sum of KS eigenvalues and the total energy, respectively. 

We used the Exciting package \cite{Gulans2014exciting} for electronic-structure calculations with the \gw\ and PBEsol$-1/2$ schemes. This package uses the full-potential linearized augmented plane wave (FPLAPW) method \cite{singh2006planewaves, Karsai2017On} to solve the Kohn-Sham single-particle equations. In the FPLAPW method, the crystal unit cell is divided into atomic spheres and interstitial regions. In the atomic spheres, atomic orbitals and spherical harmonics are used for the expansion of the KS wave functions, while in the interstitial area, plane wave basis functions are adapted. In the DFT$-1/2$ method, a half-electron is removed from anion atoms to simulate a hole excitation in the system, required for estimation of the QP self-energy effects. The corresponding self-energy potential should be trimmed with a CUT radius parameter to prevent spurious interaction between neighboring charged anions. The preferred CUT parameter is the one that maximizes the band gap of the system. Here, all ground-state calculations were performed using PBEsol functional. The size of $\mathbf{k}$- and $\mathbf{q}$-point sampling grids, the number of empty states, the number of frequencies, and also the values of basis set cut-off $R_{MT}|\mathbf{G}+\mathbf{k}|_{max}$ for all materials are presented in the supplementary information (SI). We also used the FPLAPW scheme for electronic-structure calculations using the mBJ functional, as it is implemented in the WIEN2K computer package \cite{Blaha2018WIEN2k,Blaha2020WIEN2k}. Similar computational parameters to those employed for the Exciting package were used for mBJ band-gap calculations in WIEN2K.

For DFT+$U$ calculations, we used the AFLOW$\pi$ package \cite{SUPKA2017AFLOW}, which implements the ACBN0 approach using a projection on atomic orbitals \cite{Agapito2013Effective,Agapito2016Accurate,Agapito2016Accuratetight} based on the output from Quantum ESPRESSO \cite{Giannozzi2009QUANTUM,Giannozzi2017Advanced} plane-wave code. Optimized norm-conserving (NC) Vanderbilt pseudopotentials \cite{Hamann2013Optimized} obtained from the PseudoDojo repository \cite{Setten2018The} were employed. For Li$_{2}$O, we used norm-conserving  pseudopotential obtained from the Pslibrary \cite{Andrea2014Pseudopotentials}. In these calculations, a high plane-wave energy cutoff of 120 Ry together with a dense Monkhorst-Pack mesh was utilized to ensure good convergence of all quantities (see SI for the {\bf k}-point sampling for each material in the data set. The self-consistent Hubbard $U$ values obtained within the ACBN0 scheme are presented in Table~S6.

\begin{figure}[t]
    \centering
    \includegraphics[scale=0.68]{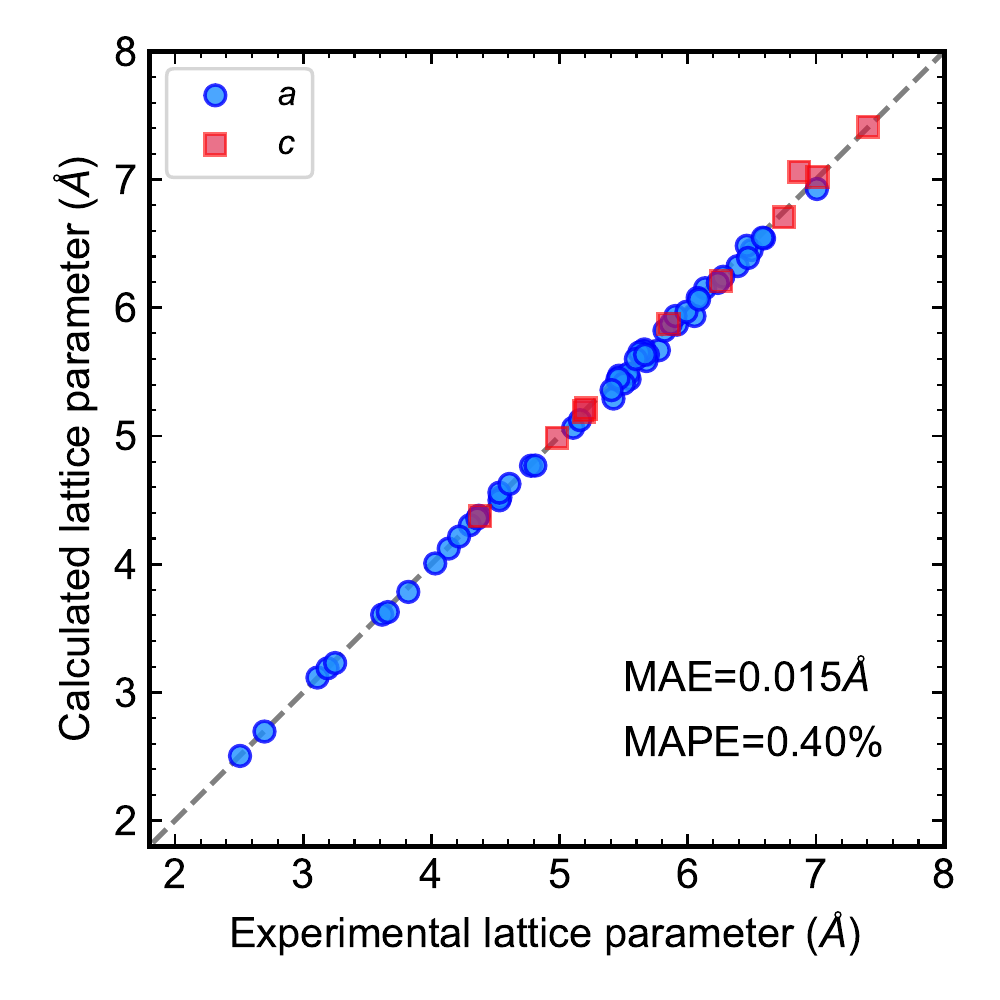}
    \caption{
             Comparison between experimental and calculated lattice parameters within PBEsol. In the case of non-cubic structures, the vertical lattice parameter $c$ is also displayed. The obtained mean absolute (percentage) error MA(P)E is also given in the figure.
            } 
 \label{fig:lattice_exp_calc} 
\end{figure}

\section*{Results and discussions}
Our data set contains 114 binary compounds with ten different crystal structures. After a broad literature survey, we separated 56 compounds from our data set, for which reliable experimental band gaps and lattice parameters are available in the literature. We note that only compounds with a non-zero PBEsol (or PBE) gap are included in the data set. 
PBEsol is found to be the most accurate GGA functional for obtaining the equilibrium parameter of crystals \cite{Schimka2011Improved,Zhang2018Performance,Haas2009Calculation}. Hence, we used the PBEsol functional to calculate the relaxed geometry of the materials in our database. The obtained equilibrium parameters are compared with the experimental data in Fig.~\ref{fig:lattice_exp_calc}, and the corresponding relative errors are given in Supplementary Table~S2. The references of the experimental data are also given in this table. As can be seen, there is an excellent agreement between the calculated and the measured lattice parameters, with a mean absolute percentage error of about 0.4\%. Hence we used the PBEsol equilibrium parameters for our band gap calculations within other desired schemes.

\begin{figure*}[!ht]
    \centering
    \includegraphics[scale=0.90]{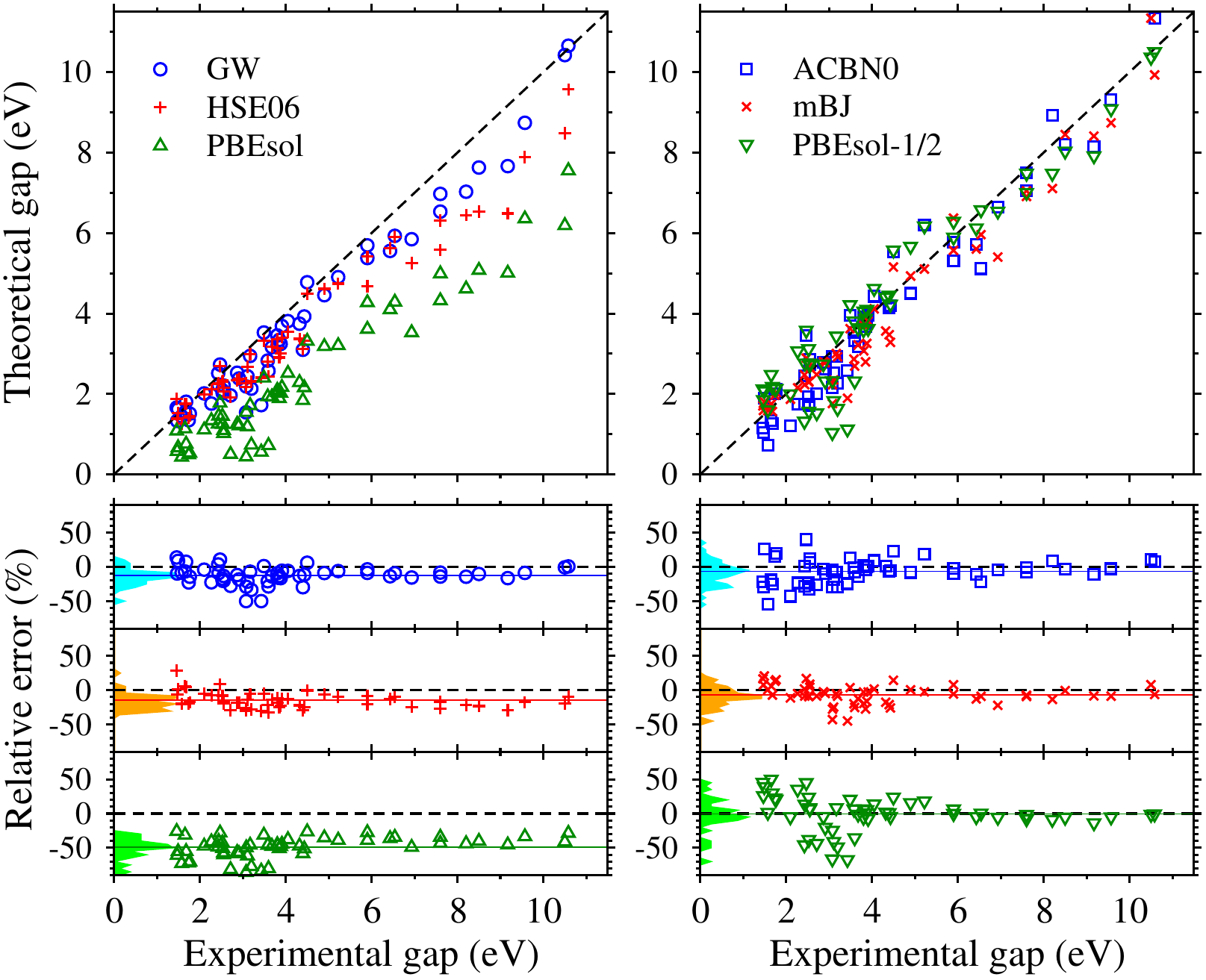}
    \caption{ 
             Comparison of theoretical and experimental band gaps for 56 compounds in our data set. The dashed lines correspond to the coincidence of theory and experiment. The relative difference between the theoretical and experimental gaps is also calculated and plotted as a function of the experimental gap. In the relative error plots, the solid lines indicate the MPE values (Table~\ref{tab:statistics}) and the shaded areas show the distribution histogram of the relative errors in arbitrary units.
            } 
 \label{fig:gaps_exp_methods} 
\end{figure*}

As others showed \cite{Karsai2018Electron,miglio2020predominance,Engel2022Zero}, zero-point vibrations \cite{mahan2013many} have a nontrivial contribution to the measured band gaps, even at extremely low temperatures. Therefore, a new theoretical scheme, zero-point renormalization (ZPR), is developed to correct the experimental band gaps for more accurate comparison with computational data. This scheme is out of the current paper's scope. However, we applied the available ZPR corrections in the literature to our reference experimental band gaps (Fig.~\ref{fig:histogram_exp} and Supplementary Table~S4). The band gaps calculated with different functionals are compared with the available experimental data in Fig.~\ref{fig:gaps_exp_methods}, while complementary details are presented in SI (Table~S5).
For more information, the direct gap of the samples with an indirect gap is also reported in this table. In these cases, the indirect band gaps were adapted for our following analysis.

The plots (Fig.~\ref{fig:gaps_exp_methods}) reveal that the band gaps obtained within the HSE06 and \gw\ methods, compared with PBEsol, are considerably improved, although they are still slightly underestimated with respect to the measured data. For better understanding, we performed a linear $y=ax+b$ least-square fit to these plots and presented the obtained slope $a$ and intercept $b$ parameters in Table~\ref{tab:statistics}. The obtained slopes for HSE06 and \gw\ are 0.767 and 0.917\,eV, respectively, while the corresponding intercepts are 0.276 and -0.165\,eV. The smaller slope and high positive intercept of HSE06, which has enhanced after the inclusion of the large gap systems, evidence the weaker performance of this scheme for large band gap semiconductors compared with \gw. The obtained error distributions (Fig.~\ref{fig:gaps_exp_methods} and Fig.~S1) show that larger band gaps are better treated within \gw, while HSE06 exhibits slightly better performance in the smaller band gap materials. These figures show that in contrast to the HSE06 and \gw\ methods, which typically exhibit a small negative systematic error, ACBN0, PBEsol$-1/2$, and to a lower extent mBJ produce scattered band gaps around the experimental values. This behavior may be attributed to the case-dependent parameters in these computationally cheaper methods, which need proper optimization. In contrast, the fraction of the exact exchange in HSE06 is fixed based on first-principles considerations \cite{Langreth1975The,Perdew1996Rationale}, while the single range-separation parameter is fitted to improve band gap prediction accuracy.

To better evaluate the performance of the functionals in comparison with the reference (experimental) band gaps, we evaluate various error statistics \cite{Civalleri2012On}, namely, 
Pearson's correlation coefficient ($r$),
Kendall's rank correlation coefficient ($\tau$) \cite{Kendall1938ANew},
the mean absolute error MAE$=\Sigma_{i}^{n}|y_{i}-y_{i,exp}|/n$,
the mean error ME$=\Sigma_{i}^{n}(y_{i}-y_{i,exp})/n$,
the error standard deviation STD$=({\Sigma_{i}^{n}(y_{i}-y_{i,exp}-ME)^2/n})^{1/2}$,
the median error MnE,
the interquartile range IQR (difference between 3rd and 1st quartile),
the median of the absolute deviations from the median MADM,
the mean absolute percentage error MAPE$=100\times \Sigma_{i}^{n}|y_{i}-y_{i,exp}|/(n y_{i,exp})$,
and the mean percentage error MPE$=100\times \Sigma_{i}^{n}(y_{i}-y_{i,exp})/(n y_{i,exp})$.
The obtained error parameters are presented in Table~\ref{tab:statistics}.

\begin{table*}[!ht]
 \caption{
          Statistical error measures of the calculated band gaps within different computational schemes, with respect to the experimental data (Exp.), \gw\ data (\gv), and ACBN0 data. The experimental data cover 56 binary compounds, while theoretical \gv\ and ACBN0 data are calculated for the whole data set of 114 compounds. The highlighted blue and black numbers show the best and the second-best records within each parameter and reference, respectively, selected with a slight tolerance (0.01\,eV or 0.1\%).
         }\label{tab:statistics}
\begin{ruledtabular}
\newcommand{\bb}{\bf\textcolor{blue}}
\newcommand{\dha}{PBEsol-1/2}
\begin{tabular}{llccccccccccccc}
 Ref.  & {}     & Pearson   &  Kendall  &   $a$     &     $b$    &   ME  &  MPE  &     STD   &     MAE   &     MAPE  &  MnE   &     IQR   &    MADM   &    MaxAE  \\
 {}   & {}     &    (\%)   &    (\%)   &           &    (eV)    &  (eV) &  (\%) &    (eV)   &     (eV)  &     (\%)  & (eV)   &    (eV)   &    (eV)   &     (eV)  \\
\midrule                                                                                                                                                        
Exp.  & PBEsol &     96.7  &     74.1  &     0.66  &     -0.49  & -1.99 & -48.8 &     1.07  &     1.99  &     48.8  &  -1.73 &     1.27  &     0.57  &     6.29  \\
      & \dha   &     95.9  &     73.2  &     0.90  &     ~0.27  & -0.15 &  -0.6 &     0.80  &     0.60  &     18.3  &  ~0.01 &     0.90  &     0.44  &     2.31  \\
      & HSE06  & \bb{98.4} & \bb{86.5} &     0.77  &     ~0.28  & -0.75 & -14.7 &     0.76  &     0.78  &     16.4  &  -0.52 &     0.81  &     0.35  &     3.88  \\
      & mBJ    &     97.8  &     83.4  &     0.90  & \bb{~0.07} & -0.36 &  -6.8 &     0.61  & {\bf0.52} & \bb{12.1} &  -0.22 &     0.81  &     0.42  &     2.58  \\
      & ACBN0  & {\bf98.1} & {\bf85.8} & \bb{1.01} &     -0.22  & -0.19 &  -6.5 & {\bf0.57} & \bb{0.48} & {\bf14.3} &  -0.26 & {\bf0.65} & {\bf0.33} & \bb{1.43} \\
      & \gv\   & \bb{98.5} &     84.1  & {\bf0.92} & {\bf-0.16} & -0.53 & -12.5 & \bb{0.50} &     0.57  & {\bf14.4} &  -0.51 & \bb{0.55} & \bb{0.29} & {\bf2.08} \\
\midrule                                                                                                                                                                          
\gv   & PBEsol & {\bf98.0} &     83.2  &     0.72  &     -0.36  & -1.30 & -42.1 &     0.70  &     1.30  &     42.1  &  -1.25 &     0.65  &     0.34  &     4.24  \\
      & \dha   &     95.8  &     78.0  & \bb{0.98} &     ~0.52  & ~0.43 &  26.1 &     0.65  &     0.67  &     35.2  &  ~0.55 &     0.72  &     0.35  &     2.69  \\
      & ACBN0  &     97.3  &     83.1  & {\bf1.07} & \bb{~0.07} & ~0.30 &  12.4 & {\bf0.58} &     0.53  &     23.0  &  ~0.33 &     0.77  &     0.37  & {\bf1.99} \\
      & mBJ    & {\bf98.1} & {\bf86.5} & \bb{0.98} & {\bf~0.29} & ~0.23 &  12.7 & \bb{0.44} & {\bf0.35} & {\bf15.8} &  ~0.17 &     0.36  & \bb{0.17} &     2.13  \\
      & HSE06  & \bb{99.3} & \bb{89.3} &     0.84  &     ~0.44  & -0.11 &   4.4 & \bb{0.43} & \bb{0.30} & \bb{12.4} &  -0.04 &     0.43  & {\bf0.23} & \bb{1.95} \\
\midrule                                                                                                                                                                          
ACBN0 & PBEsol &     94.7  &     73.7  &     0.64  & {\bf-0.26} & -1.60 & -44.3 &     1.04  &     1.60  &     44.3  &  -1.57 &     1.25  &     0.56  &     5.46  \\
      & \dha   &     95.0  &     74.6  & \bb{0.88} &     ~0.57  & ~0.13 &  15.3 &     0.77  &     0.62  &     29.7  &  ~0.16 &     1.05  &     0.51  & \bb{1.69} \\
      & HSE06  & {\bf96.4} &     81.2  & {\bf0.74} &     ~0.54  & -0.41 &  -1.3 &     0.81  &     0.69  & {\bf22.6} &  -0.36 &     1.08  &     0.54  &     3.17  \\
      & mBJ    & \bb{97.2} & {\bf82.5} & \bb{0.89} &     ~0.35  & -0.07 &   5.2 & {\bf0.60} & \bb{0.49} & \bb{19.3} &  -0.05 & {\bf0.94} & {\bf0.47} & {\bf1.83} \\
      & \gv\   & \bb{97.3} & \bb{83.1} & \bb{0.89} & \bb{~0.11} & -0.30 &  -3.7 & \bb{0.58} & {\bf0.53} & \bb{19.2} &  -0.33 & \bb{0.77} & \bb{0.37} &     1.99  \\
\end{tabular}
\end{ruledtabular}
\end{table*}

The inexpensive PBEsol$-1/2$, ACBN0, and mBJ schemes have small mean error values of -0.15, -0.19, and -0.36\,eV, while \gw\ and HSE06 give high ME values of about -0.53 and -0.75\,eV, respectively. This fact does not indicate the higher accuracy of the cheaper methods; it rather demonstrates the mentioned unsystematic error behavior of these methods compared with the more systematic errors of HSE06 and \gw. The unsystematic error distributions are usually accompanied by significant error cancellation and thus lead to low mean error values. Therefore, MAE is more appropriate for comparing the accuracy of these methods. The ACBN0 and mBJ methods exhibit lower MAE values (0.48 and 0.52\,eV) and thus are more accurate than PBEsol$-1/2$, HSE06, and \gw\ with MAE values above 0.57\,eV. In terms of MAPE, mBJ exhibits the lowest error (12.1\%), while ACBN0 and \gw\ show slightly larger errors. In addition to the classical statistical measures, we have used MnE, MADM, and IQR, which are less sensitive to outliers \cite{Kaliraj2022Big,Civalleri2012On}. The results show that PBEsol$-1/2$ and mBJ have the lowest MnE value of 0.01 and -0.22\,eV, respectively, while \gw\ has the lowest IQR and MADM values. 

Correlation coefficients are statistical parameters that evaluate the accuracy of the models in predicting the trend of the target quantities compared with the reference data. In this regard, we calculated Pearson's and Kendall's rank correlations between theoretical and experimental band gaps and presented the results in Table~\ref{tab:statistics}. \gw\ has the highest Pearson's correlation with the experiment, followed by HSE06 and then other methods. Considering Kendall's rank correlation, the mBJ, \gw, ACBN0, and HSE06 schemes give comparable values in the range of 0.83-0.87, whereas PBEsol$-1/2$ and PBEsol provide smaller values.

Among the 13 statistical parameters presented in Table~\ref{tab:statistics}, we have highlighted ten more proper overall measures for a general comparative evaluation of the selected theoretical methods. The ME and MPE parameters are not used for this evaluation because of the involved error cancellation in these parameters. The median error MnE is also sensitive to the systematic or unsystematic error behavior of the methods and hence may not be useful for our comparative evaluation. Taking into account the highlighted parameters, one can conclude that \gw\ and ACBN0 provide the best performance for predicting the experimental band gap of semiconductors. \gw\ is the most (second) accurate method according to four (four) parameters, whereas ACBN0 is the best (second) scheme within three (six) statistical measures. The mBJ and HSE06 functionals take the next places when compared with the experimental data. mBJ gives the best MAPE and $b$ parameters and the second-best MAE value, while HSE06 exhibits the highest Pearson and Kendall correlations. In fact, mBJ is more accurate than HSE06 in the reproduction of the measured band gaps, in agreement with recent publications \cite{Rauch2020Accurate,Tran2021Bandgap}. Taking into account the expensive computational cost of HSE06 and \gw, the above discussion indicates that ACBN0 and mBJ are clearly more efficient than HSE06 for band gap calculations. Moreover, ACBN0 is found to be a very good and reliable replacement instead of \gw\ for high-throughput screening of band gaps.

\begin{figure*}[!th]
    \centering
    \includegraphics[scale=0.65]{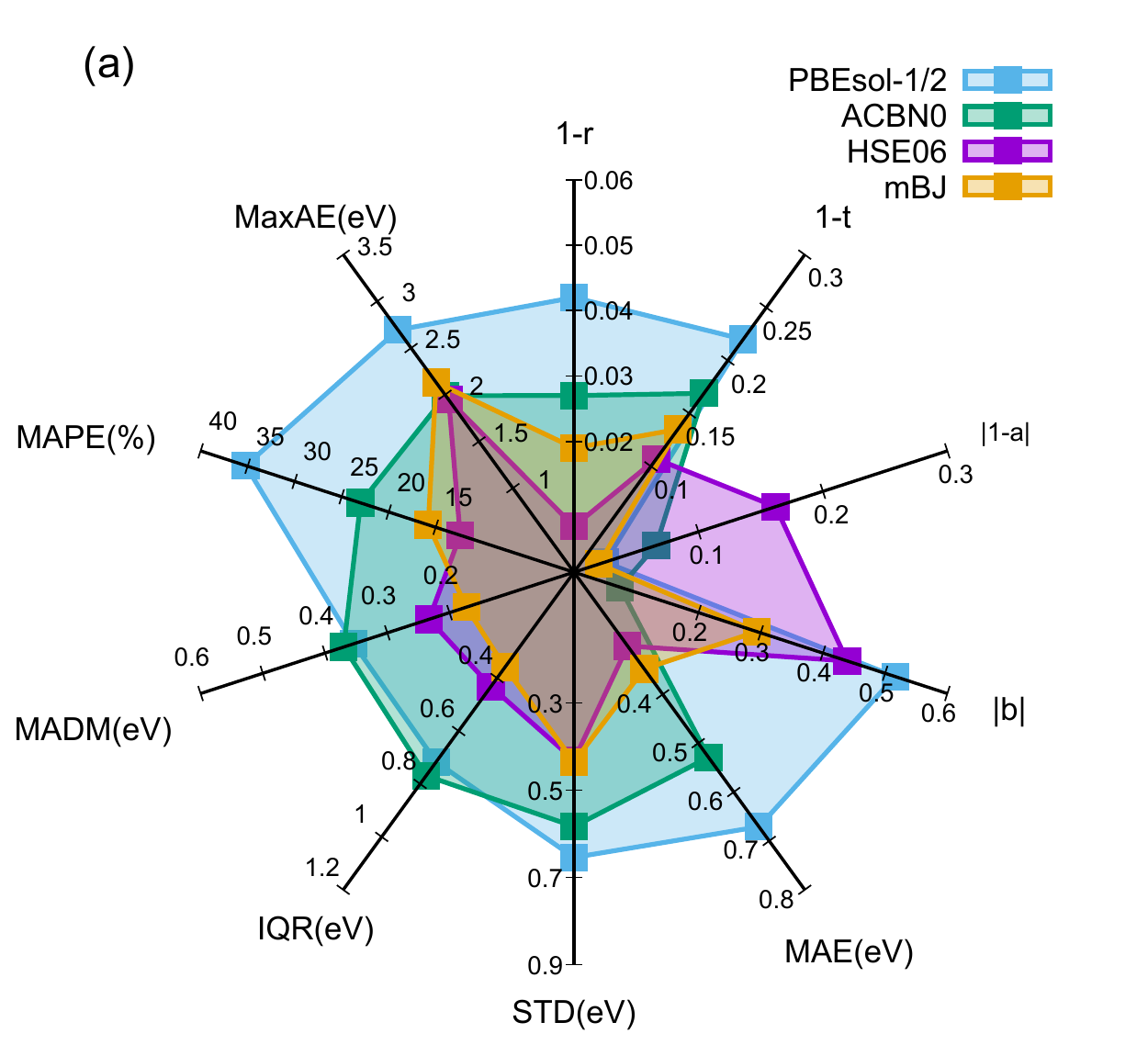}
    \includegraphics[scale=0.65]{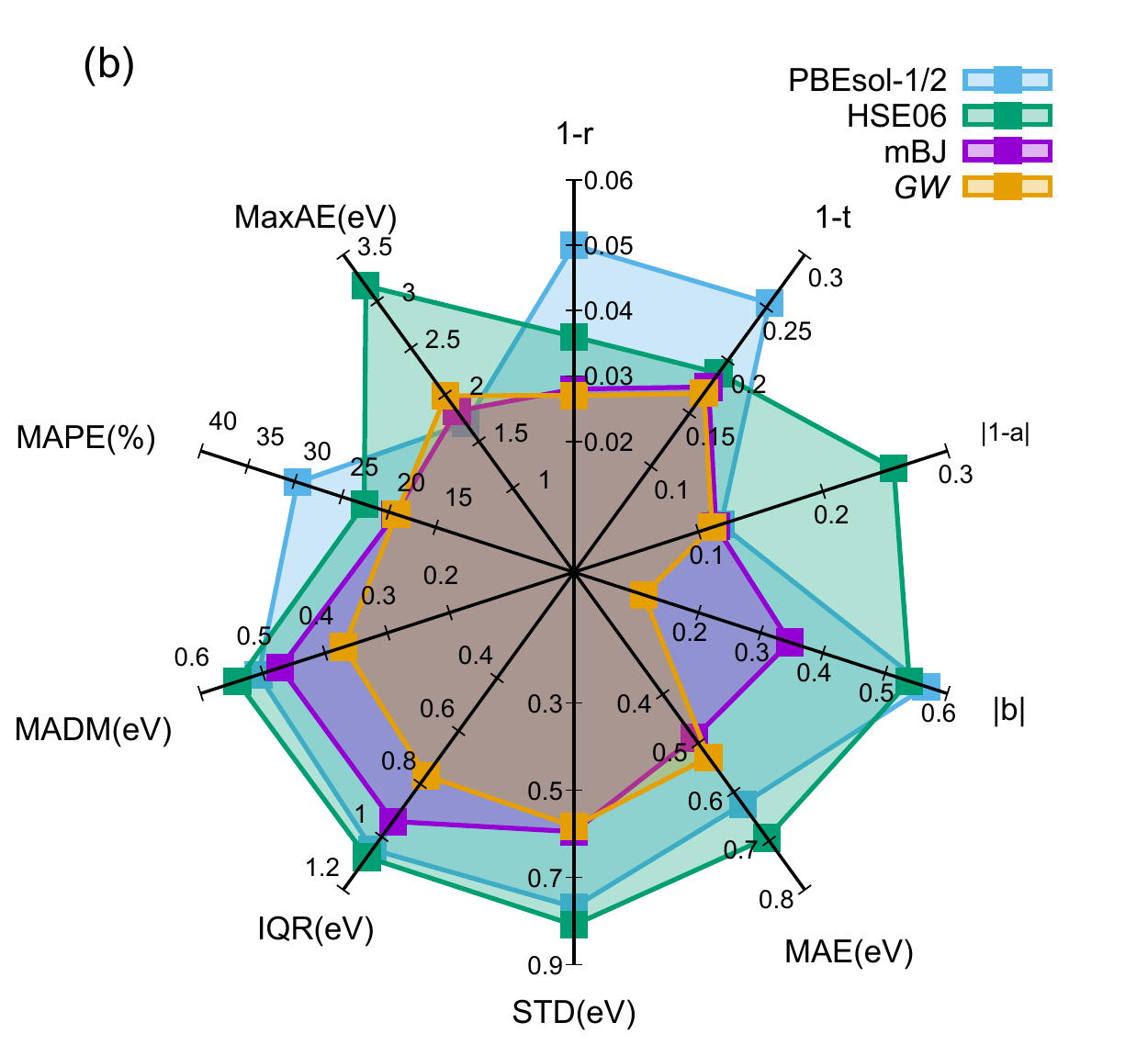}
    \caption{
             Radar plots showing the ten more important statistical error quantities (Table~\ref{tab:statistics}) calculated for 114 binary materials within different theoretical schemes with reference to the (a) \gw\ and (b) ACBN0 band gap data.
            } 
 \label{fig:radar} 
\end{figure*}

Based on the above analysis, in the rest of the paper, we choose \gw\ and ACBN0 as our first-principles reference methods and evaluate the performance of the other approaches in the whole data set of 114 binary materials. The calculated statistical measures compared with \gw\ and ACBN0 are separately given in Table~\ref{tab:statistics}, and the corresponding radar plots and histograms are displayed in Fig.~\ref{fig:radar} and Figs.~S2 and S3 (SI), respectively.

In comparison with the \gw\ reference data, HSE06 exhibits the best agreement with six first and one second records. The mBJ scheme takes second place with three first and five second records, and the ACBN0 takes third place with only one first and three second records. The weaker performance of ACBN0 in reproducing the \gv\ band gaps, compared with mBJ, looks strange because of the very similar performance of ACBN0 and \gv\ in reproducing the experimental band gaps. A more accurate inspection of the error histograms (Fig.~\ref{fig:gaps_exp_methods}) and ME values (Table~\ref{tab:statistics}) indicates that, as it was mentioned before, mBJ lies between \gv\ and ACBN0 from the point of view of the underestimation of the experimental gap. As a result, mBJ is more successful than ACBN0 in reproducing the \gw\ band gaps.

On the other hand, adapting the ACBN0 band gaps as our reference data, we found that \gw\ and mBJ exhibit the best performances (Table~\ref{tab:statistics} and Fig.~\ref{fig:radar}). Although \gv\ has eight first records, mBJ has a slightly lower MAE and hence really competes with \gv\ in reproducing the ACBN0 band gaps. HSE06 take third place only with three second records. The point that seems strange is the fact that while HSE06 is the first and ACBN0 is the third successful method in producing the \gv\ band gaps (mentioned in the previous paragraph), in reproducing the ACBN0 band gaps, \gv\ is the most prosperous scheme. Again, we attribute this behavior to the rather systematic error behavior of \gv\ and HSE06 and the unsystematic error behavior of ACBN0. The obtained ME values indicate that the \gv\ band gaps are statistically distributed above the HSE06 and below the ACBN0 data, being closer to the HSE06 data. For example, compared with the experimental references, the ME value for HSE06, \gv, and ACBN0 are -0.75, -0.53, and -0.19\,eV, respectively, which confirm the claimed order between these three schemes. The same arguments apply to theoretical references as well. As a result, compared with the \gv\ reference, HSE06 is more consistent than ACBN0, while in comparison with the ACBN0 data, the \gv\ scheme is more accurate than HSE06.

\begin{figure*}[!ht]
    \centering
    \includegraphics[scale=0.5]{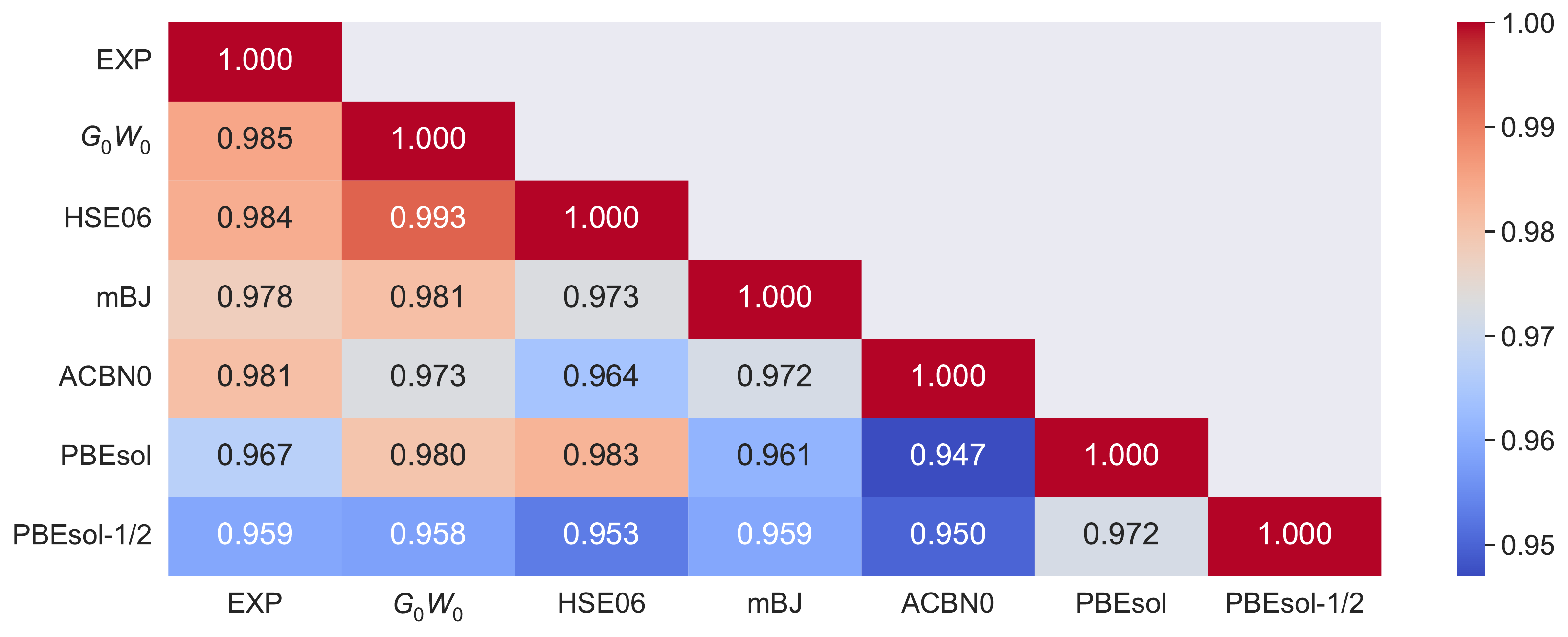}
    \caption{
             Obtained Pearson's correlation heatmap among the selected theoretical schemes and experimental (Exp) data.
            } 
 \label{fig:corr_heatmap} 
\end{figure*}

Finally, we calculate the Pearson coefficient to analyze the mutual correlations between the selected theoretical schemes and the experiment. These results may be useful to identify the best low-fidelity band gaps for machine learning modeling of the high-fidelity \gw\ band gaps. The obtained Pearson's correlation map for the considered approximations is presented in Fig.~\ref{fig:corr_heatmap}. It is clear that all theoretical methods correlate quite well (96\% and more) with the experiment. Hence, all selected schemes, even conventional PBEsol, are expected to provide a reliable description of the band gap trends in different situations. Among the economic schemes, mBJ exhibits the highest correlation with the \gw\ method. Taking into account this fact along with the high agreement between the mBJ band gaps and the \gv\ ones, we conclude that the mBJ band gap is a very good descriptor for machine learning prediction of the quasiparticle \gv\ band gaps.

\section*{Conclusions}
We have performed an extensive benchmark calculation within various exchange-correlation schemes, including \gw, hybrid HSE06, GGA PBEsol, meta-GGA mBJ, quasiparticle-corrected PBEsol$-1/2$, and pseudo-hybrid self-consistent ACBN0 functionals to evaluate their performance in predicting band gaps of semiconductors. We used a data set of 56 binary materials for evaluating the selected schemes with respect to the experiment, while in a broader group of 114 binary compounds, the \gw\ and ACBN0 schemes were adapted as a reference to assess other theoretical methods. Compared with the experiment, \gw\ outperforms overall all other functionals, although the mean absolute error (MAE) within ACBN0 and mBJ (0.48 and 0.51\,eV) are slightly better than \gw\ (0.57\,eV). HSE06 shows the best correlations with the experiment, although its statistical error measures are less refined than ACBN0, especially for the larger band gap semiconductors. Overall, it is argued that  ACBN0 is a promising alternative for the expensive \gv\ scheme for high-throughput scanning of the band gap of semiconductors and insulators.

In the group of 114 compounds, HSE06 shows the best agreement with the reference \gw\ data, with a very low MAE of about 0.30\,eV and mBJ closely follow HSE06 in reproducing the reference band gaps. Despite its success in predicting the experimental data, the performance of ACBN0 in producing the \gv\ results is considerably less than mBJ. The reason is that the \gv\ and HSE06 band gaps are mainly below the experimental values, while the ACBN0 and PBEsol$-1/2$ values are scattered around the measured data, leading to the lowest mean errors of -0.19 and -0.15\,eV, respectively. The mBJ functional exhibits a median behavior with a mean error of -0.36\,eV, thus showing more agreement with \gv, compared with ACBN0. We conclude that the economic mBJ method helps produce cheap band gap descriptors for artificial intelligence modeling of the expensive \gw\ band gaps.

\section*{Acknowledgments}
This work was supported by RFBR-INSF grant number 20-53-56065.
%

\onecolumngrid
\newpage
\setcounter{section}{0}
\setcounter{equation}{0}
\setcounter{figure}{0}
\setcounter{table}{0}

\renewcommand {\thesubsection} {Supplementary Note \arabic{subsection}}
\renewcommand{\theequation}{S\arabic{equation}}
\renewcommand{\thetable}{S\arabic{table}}
\renewcommand{\thefigure}{S\arabic{figure}}
\renewcommand{\theHfigure}{A\arabic{figure}}
\renewcommand{\figurename}{Supplementary Fig}

\section*{\fontsize{14}{25}\selectfont {Supplementary Information}}
Table~\ref{Tab:table_S1} shows the most important parameters used for \gw\ calculations in the EXCITING package. These parameters include the {\bf k}-point sampling, the {\bf q}-point sampling, the number of empty states, the number of frequencies, and the values of RKmtmax. 
The experimental lattice parameters of 56 structures, compared with calculated values, are illustrated in Table~\ref{Tab:table_S2}.
In Table~\ref{Tab:table_S3}, we present the 
{\bf k}-point sampling used in Quantum-Espresso code for ACBN0 calculations, as described in the main text. 
The obtained ACBN0 band gaps and Hubbard $U$ parameters of binary materials are tabulated in Tables~\ref{Tab:dataset} and \ref{Tab:Hubbard_U}, respectively.
In Table~\ref{Tab:dataset}, the list of all binary compounds with their calculated band gaps in the functionals considered in this benchmark study is presented. This table also contains the Materials Project identification number (MP id), the number of space groups, the experimental band gap if it is available, and also the type of PBEsol band gap.
Table~\ref{Tab:zpr_values} gives the collected available zero-phonon renormalization (ZPR) values in the literature.
{
\makeatletter
\renewcommand\table@hook{\fontsize{8}{6}\selectfont}
\makeatother
\begin{table*}[!ht]
\centering
 \caption{
          Computational parameters used for $G_{0}W_{0}$@PBEsol calculations. From left to right: Wavevector sampling for the electronic states; wavevector sampling for the phonon states; the number of empty states; the number of frequencies; the basis set cut-off $R_{MT}|\mathbf{G}+\mathbf{k}|_{max}$.
         }\label{Tab:table_S1}
\begin{ruledtabular}
\begin{tabular}{lC{1.5cm}C{1.5cm}C{1cm}cc|lC{1.5cm}C{1.5cm}C{1cm}cc}
{} & {\bf k}-point sampling & {\bf q}-point sampling &  \# of empty states &  frequency &  rgkmax & {} & {\bf k}-point sampling & {\bf q}-point sampling &  \# of empty states &  frequency &  rgkmax \\
\midrule
AgBr-216  &   $16\times16\times16$ &       $8\times8\times8$ &     16 &      32 &     7.5  &     GaP-186   &    $14\times14\times7 $ &       $8\times8\times4$ &     18 &      32 &     7.0 \\
AgBr-225  &   $16\times16\times16$ &       $8\times8\times8$ &     12 &      32 &     7.5  &     GaP-216   &    $16\times16\times16$ &       $6\times6\times6$ &     16 &      32 &     8.0 \\
AgCl-221  &   $16\times16\times16$ &       $8\times8\times8$ &     16 &      32 &     7.0  &     GeC-186   &    $14\times14\times7 $ &       $8\times8\times4$ &     14 &      18 &     7.0 \\
AgCl-225  &   $16\times16\times16$ &       $8\times8\times8$ &     16 &      32 &     7.5  &     GeC-216   &    $12\times12\times12$ &       $8\times8\times8$ &     16 &      32 &     7.0 \\
AgI-216   &   $14\times14\times14$ &       $8\times8\times8$ &     18 &      32 &     7.5  &     InP-216   &    $16\times16\times16$ &       $8\times8\times8$ &     16 &      32 &     7.0 \\
AgI-225   &   $16\times16\times16$ &       $8\times8\times8$ &     18 &      32 &     8.0  &     K2O-225   &    $16\times16\times16$ &       $8\times8\times8$ &     14 &      32 &     7.5 \\
AlAs-186  &   $14\times14\times8 $ &       $8\times8\times4$ &     18 &      32 &     7.0  &     K2S-225   &    $15\times15\times15$ &       $6\times6\times6$ &     18 &      32 &     7.5 \\
\bottomrule
\end{tabular}
\end{ruledtabular}
\end{table*}
}
{
\makeatletter
\renewcommand\table@hook{\fontsize{8}{6}\selectfont}
\makeatother
\begin{table*}[!ht]
\centering
 \caption{
          Theoretical and experimental lattice constants (\AA) for the set of 56 binary compounds. Theoretical lattice parameters are calculated by PBEsol functional.
         }\label{Tab:table_S2}
\begin{ruledtabular}
\begin{tabular}{lrrrrrrr}
{} &  exp\_a &  exp\_c &  conventional\_a &      c &  relative\_error\_conv\_a &  relative\_error\_c & ref\\
\midrule
AgBr-225 &  5.775 &        &           5.669 &        &                  1.839 &                   & \cite{madelung2004semiconductors}\\
AgCl-225 &  5.546 &        &           5.443 &        &                  1.854 &                   & \cite{Zoroddu2001First}\\
AgI-216  &  6.499 &        &           6.451 &        &                  0.745 &                   & \cite{madelung2004semiconductors}\\
AlAs-216 &  5.661 &        &           5.675 &        &                  0.244 &                   & \cite{madelung2004semiconductors}\\
AlN-186  &  3.111 &  4.979 &           3.115 &  4.986 &                  0.139 &             0.152 & \cite{madelung2004semiconductors}\\
AlN-216  &  4.370 &        &           4.377 &        &                  0.171 &                   & \cite{Vurgaftman2003Band}\\
AlP-216  &  5.464 &        &           5.470 &        &                  0.125 &                   & \cite{madelung2004semiconductors}\\
\bottomrule
\end{tabular}
\end{ruledtabular}
\end{table*}
}
{
\makeatletter
\renewcommand\table@hook{\fontsize{8}{6}\selectfont}
\makeatother
\begin{table*}[!ht]
\centering
 \caption{
          {\bf k}-point sampling used for ACBN0 calculations
         }\label{Tab:table_S3}
\begin{tabular}{llll}
\toprule
\toprule
{}~~~~~~~~~~~~~~~~~~~~~~~~~~~~~~~~~~ & {\bf k}-point sampling~~~~~~~~~~~~~~~~~~~~~~~~~~~~~~~~~~ & {}~~~~~~~~~~~~~~~~~~~~~~~~~~~~~~~~~~ & {\bf k}-point sampling \\
\midrule
AgBr-216     & $ 12 \times 12 \times 12  $         & GaP-186      & $ 13 \times 13 \times 7   $ \\
AgBr-225     & $ 13 \times 13 \times 13  $         & GaP-216      & $ 14 \times 14 \times 14  $ \\
AgCl-221     & $ 13 \times 13 \times 13  $         & GeC-186      & $ 16 \times 16 \times 8   $ \\
AgCl-225     & $ 14 \times 14 \times 14  $         & GeC-216      & $ 16 \times 16 \times 16  $ \\
AgI-216      & $ 12 \times 12 \times 12  $         & InP-216      & $ 13 \times 13 \times 13  $ \\
AgI-225      & $ 13 \times 13 \times 13  $         & K2O-225      & $ 12 \times 12 \times 12  $ \\
AlAs-186     & $ 13 \times 13 \times 7   $         & K2S-225      & $ 10 \times 10 \times 10  $ \\
\bottomrule
\end{tabular}
\end{table*}
}
\begin{table*}[!ht]
\centering
 \caption{
          Available values of ZPR for materials used in this study. The values are obtained from references \cite{Chen2018Nonempirical} and \cite{Cardona2005Isotope}.
         }\label{Tab:zpr_values}
\begin{tabular}{lrr}
\toprule
\midrule
{} &    ~~~~~~~~~~~~~~~~~~~~~~ZPR (eV) \\
\midrule
AlAs-216 &  0.039 \\
AlN-186  &  0.239 \\
AlP-216  &  0.023 \\
AlSb-216 &  0.039 \\
BN-216   &  0.340 \\
CdS-186  &  0.068 \\
CdSe-186 &  0.034 \\
CdTe-216 &  0.017 \\
CuBr-216 &  0.005 \\
CuCl-216 &  0.031 \\
CuI-216  & -0.009 \\
GaAs-216 &  0.054 \\
GaN-186  &  0.173 \\
GaP-216  &  0.085 \\
InP-216  &  0.048 \\
LiCl-225 &  0.170 \\
LiF-225  &  1.150 \\
MgO-225  &  0.530 \\
NaCl-225 &  0.170 \\
SiC-216  &  0.110 \\
ZnO-186  &  0.156 \\
ZnS-216  &  0.080 \\
ZnSe-216 &  0.047 \\
ZnTe-216 &  0.033 \\
\bottomrule
\end{tabular}
\end{table*}

\clearpage
\setlength{\LTcapwidth}{\textwidth}
{
\makeatletter
\renewcommand\table@hook{\fontsize{9}{8}\selectfont}
\makeatother
\begin{ruledtabular}
\begin{longtable*}{llrrrrrrrrrrr}
\caption[0.5\textwidth]{
         List of the binary materials considered in the present benchmark study. From left to right: Material; Materials Project ID; available experimental band gaps; band gaps calculated with different approaches.
        }\label{Tab:dataset}\\\toprule
\endfirsthead
\caption* {\textbf{Table \ref{Tab:dataset}} \it{(Continued)}}\\\toprule
\endhead
\endfoot
\bottomrule
\endlastfoot
\toprule
Material~~~~ & MP id~~~~ & ~~~~SG & ~~~~~Expt. & ~~~~PBEsol & ~~~~type & ~~~~mBJ & ~~~~DFT$-1/2$ & ~~~~HSE06 & ~~~~ACBN0 & ~~~~$G_{0}W_{0}$ & DIR & INDIR \\
\midrule
    AgBr &  mp-866291 & 216 &                                           &       0.889 &   direct &    2.891 &        1.942 &    2.248 &      2.390   &   2.199 &    2.199 &           \\
    AgBr &   mp-23231 & 225 &    2.712\cite{madelung2004semiconductors} &       0.502 &    indir &    2.480 &        1.526 &    1.920 &      2.002   &   1.963 &    3.824 &     1.963 \\
    AgCl & mp-1060502 & 221 &                                           &       0.453 &    indir &    2.685 &        1.403 &    1.961 &      1.959   &   1.721 &    3.574 &     1.721 \\
    AgCl &   mp-22922 & 225 &    3.200\cite{madelung2004semiconductors} &       0.731 &    indir &    2.924 &        1.637 &    2.310 &      2.267   &   2.130 &    4.691 &     2.130 \\
     AgI &   mp-22925 & 216 &    2.910\cite{madelung2004semiconductors} &       1.212 &   direct &    2.775 &        2.318 &    2.376 &      2.615   &   2.382 &    2.382 &           \\
     AgI &   mp-22919 & 225 &                                           &       0.386 &    indir &    1.731 &        1.427 &    1.514 &      1.713   &   1.383 &    3.764 &     1.383 \\
    AlAs &    mp-8881 & 186 &                                           &       1.618 &    indir &    2.408 &        3.149 &    2.384 &      2.036   &   2.138 &    2.419 &     2.138 \\
\bottomrule
\end{longtable*}
\end{ruledtabular}
}

{
\makeatletter
\renewcommand\table@hook{\fontsize{8}{7}\selectfont}
\makeatother
\begin{table}
 \centering
 \caption{
         Converged values of $U_{eff}$ (in eV) obtained from ACBN0 calculations.  
        }\label{Tab:Hubbard_U}
\begin{ruledtabular}
\begin{tabular}{lll|lll}
{}        &         U1        &            U2        &  {}        &           U1          &          U2         \\
\midrule
BN-216    &    U\_B  $=$   0.24 &    U\_N   $=$     3.63  &   AgCl-221  &   U\_Ag   $=$    8.45 &   U\_Cl   $=$    3.46 \\
BN-194    &    U\_B  $=$   0.16 &    U\_N   $=$     3.88  &   AgI-225   &   U\_Ag   $=$    9.27 &    U\_I   $=$     3.3 \\
BP-216    &    U\_B  $=$   0.71 &    U\_P   $=$     0.92  &   AlAs-186  &   U\_Al   $=$    0.09 &   U\_As   $=$    2.01 \\
BAs-216   &   U\_As  $=$   0.85 &    U\_B   $=$     0.64  &   AlN-225   &   U\_Al   $=$    0.11 &    U\_N   $=$    4.93 \\
AlN-186   &   U\_Al  $=$   0.05 &    U\_N   $=$     5.26  &   AlP-186   &   U\_Al   $=$    0.08 &    U\_P   $=$    2.27 \\
AlP-216   &   U\_Al  $=$   0.08 &    U\_P   $=$      2.1  &   BaCl2-225 &   U\_Ba   $=$     0.0 &   U\_Cl   $=$    7.27 \\
AlAs-216  &   U\_Al  $=$   0.09 &   U\_As   $=$     1.85  &   BaO-129   &   U\_Ba   $=$    0.16 &    U\_O   $=$    9.54 \\
\bottomrule
\end{tabular}
\end{ruledtabular}
\end{table}
}

\begin{figure*}[!ht]
   \centering
   \includegraphics[scale=0.25]{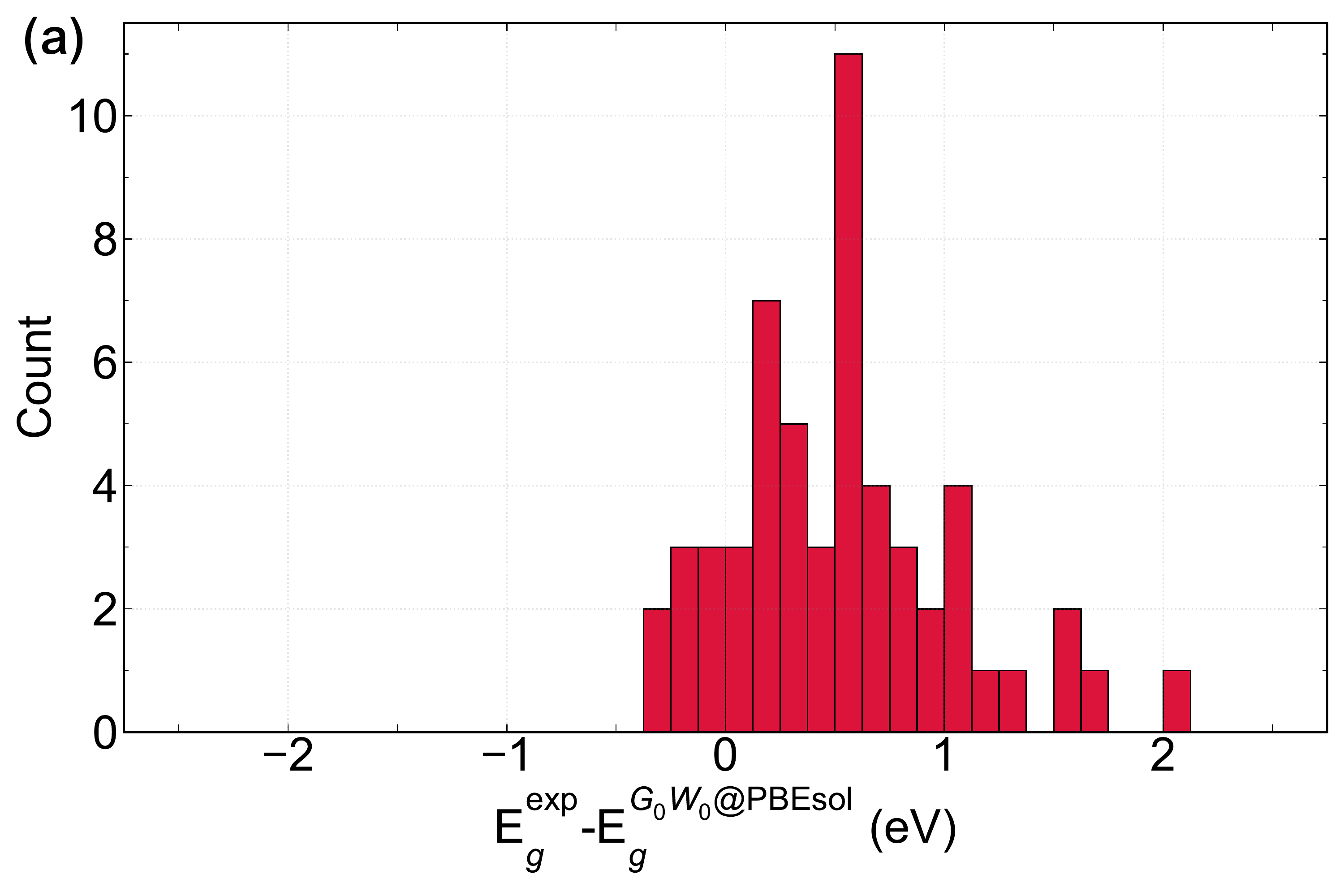}
   \includegraphics[scale=0.25]{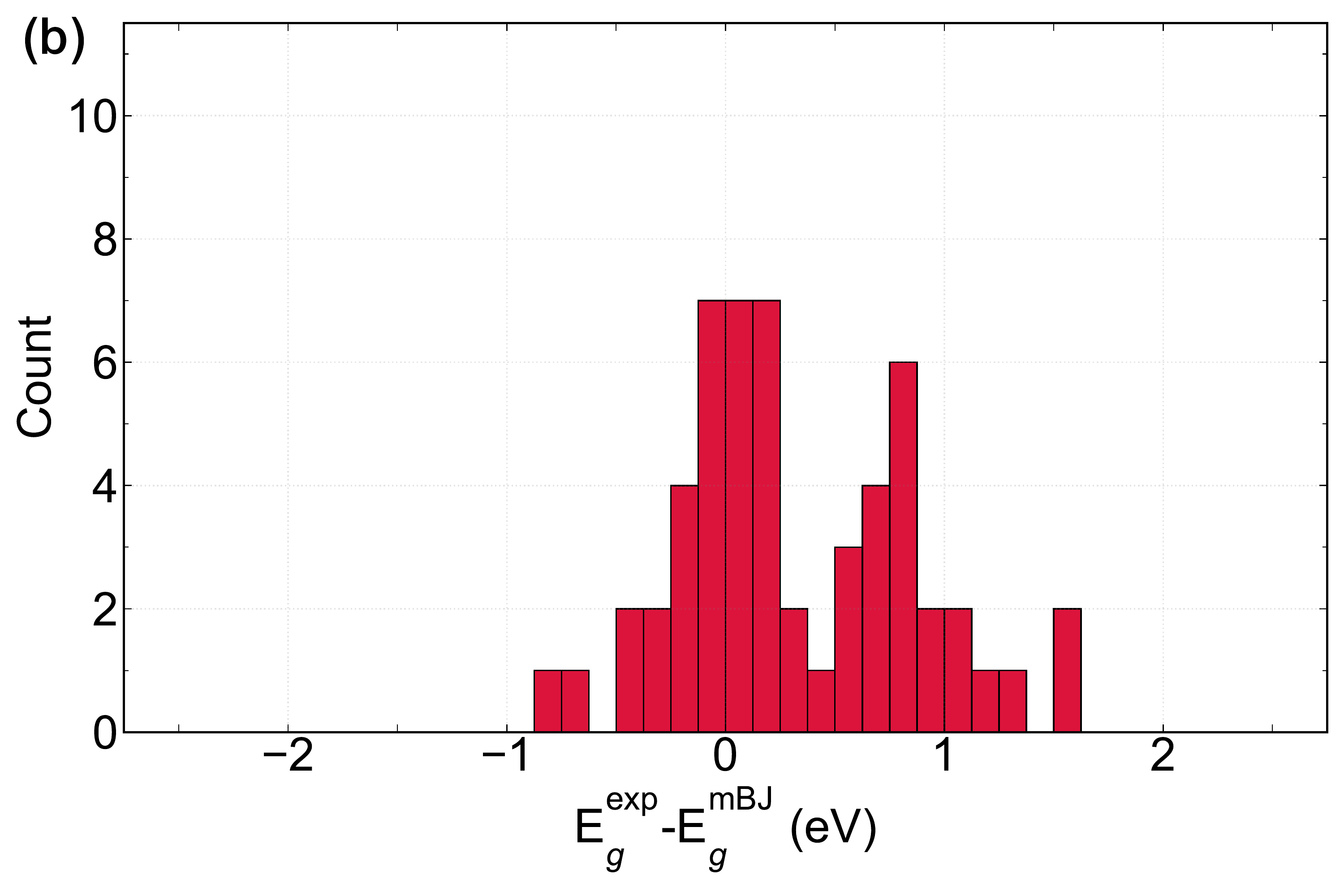} \\
   \includegraphics[scale=0.25]{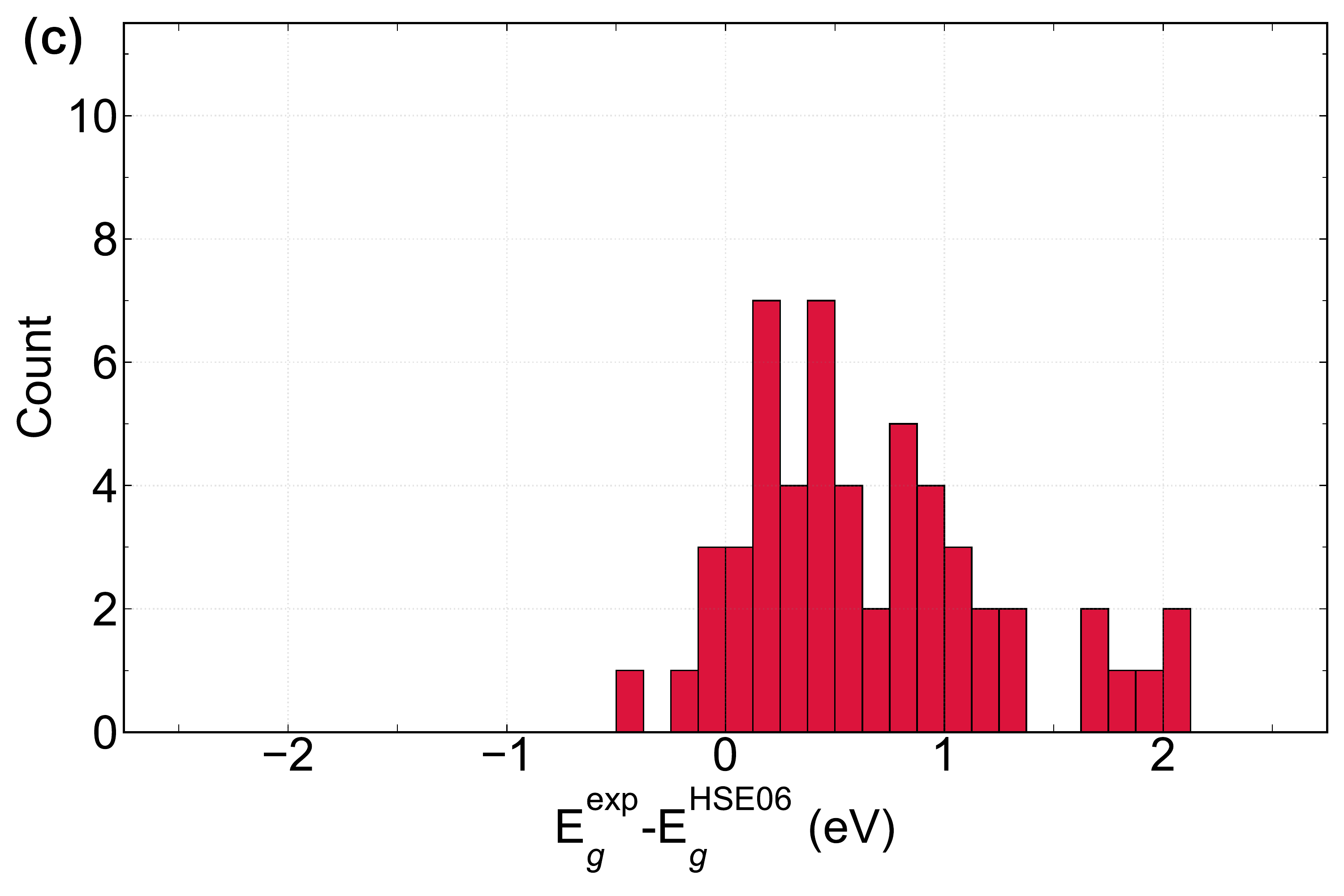}
   \includegraphics[scale=0.25]{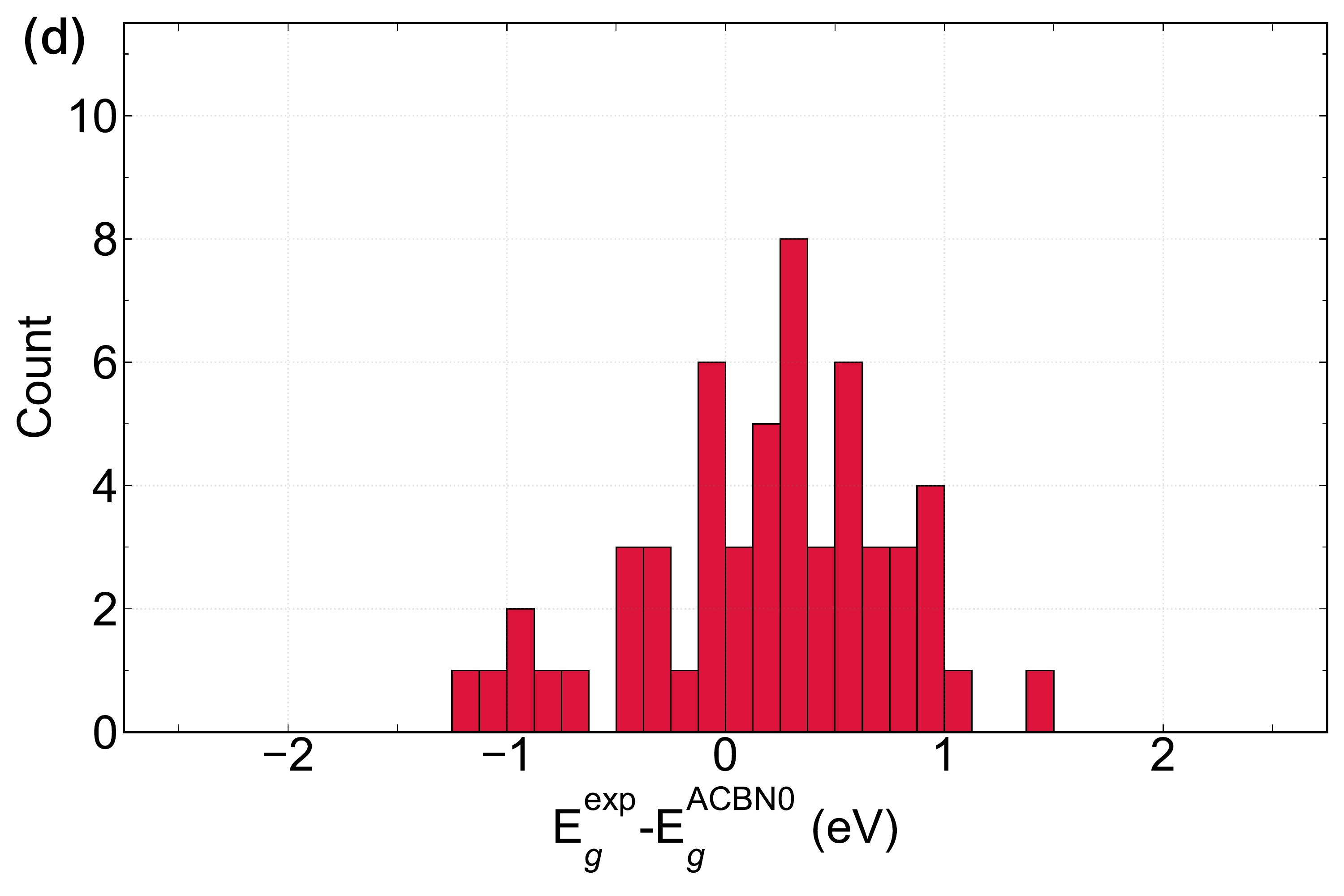} \\
   \includegraphics[scale=0.25]{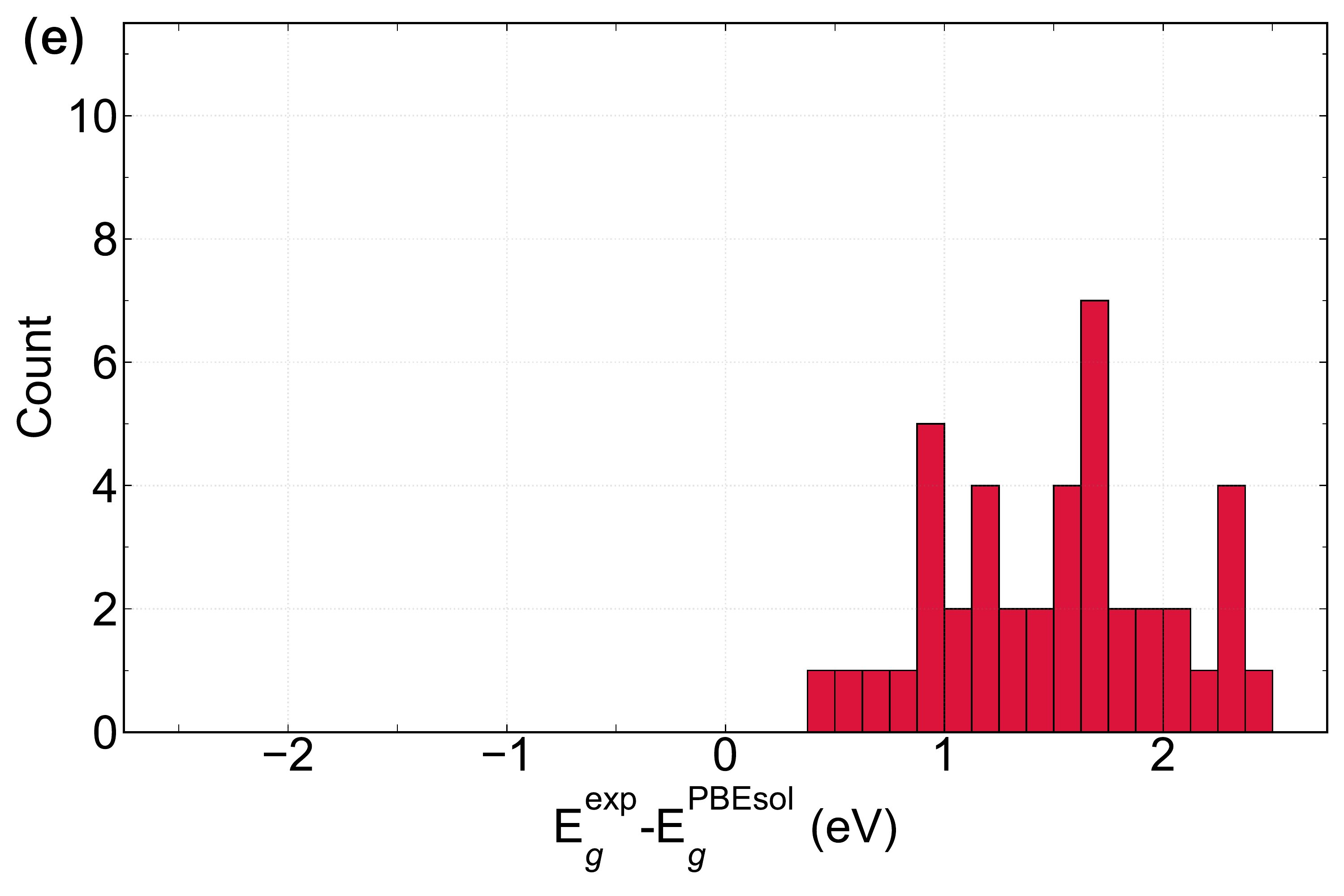}
   \includegraphics[scale=0.25]{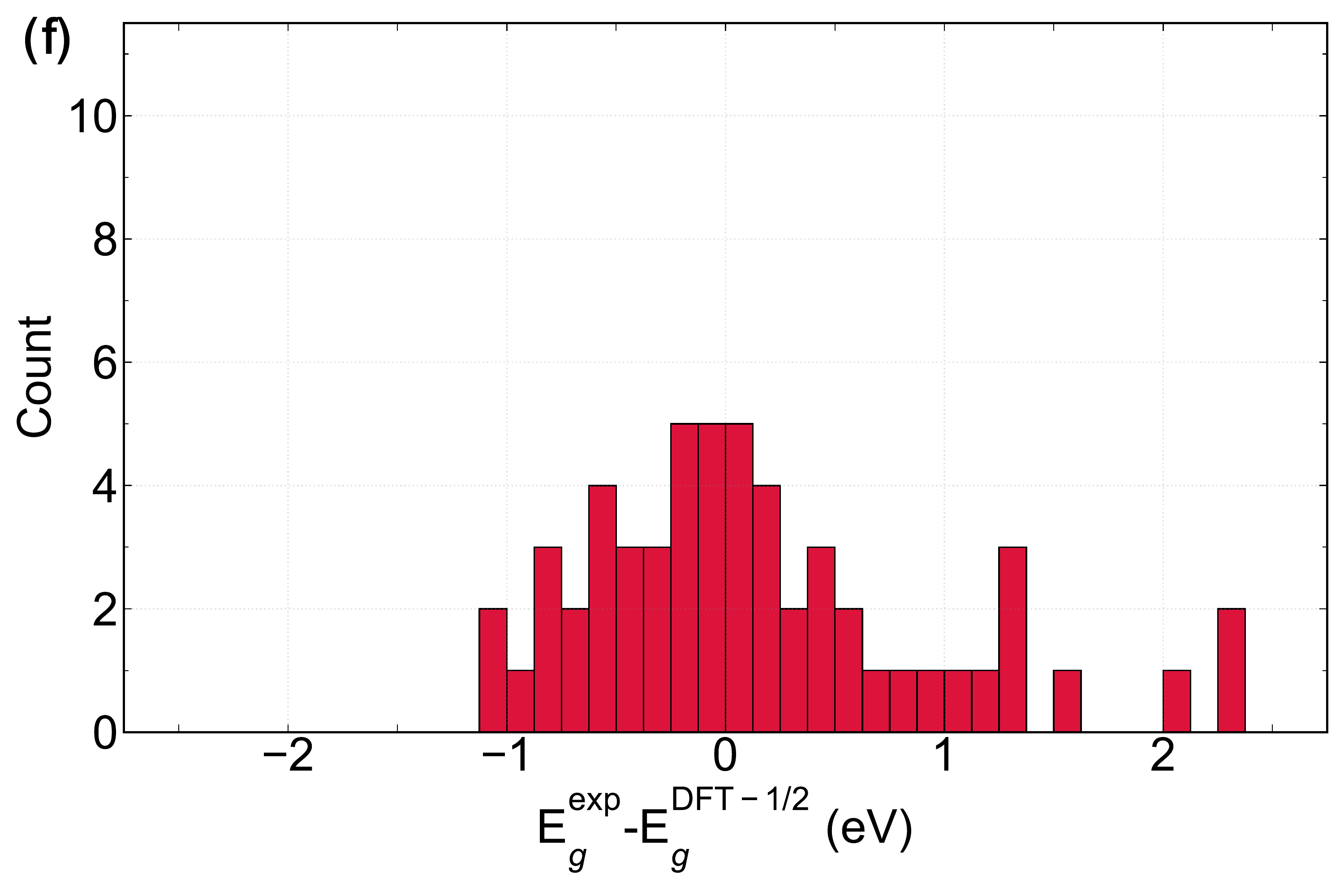} \\
   \caption{
            Distribution histogram of differences between experimental and theoretical band gaps for (a) \gw, (b) mBJ, (c) HSE06, (d) ACBN0, (e) PBEsol and (f) PBEsol$-1/2$.
           } 
\end{figure*}

\begin{figure*}[!ht]
    \centering 
    \includegraphics[scale=0.25]{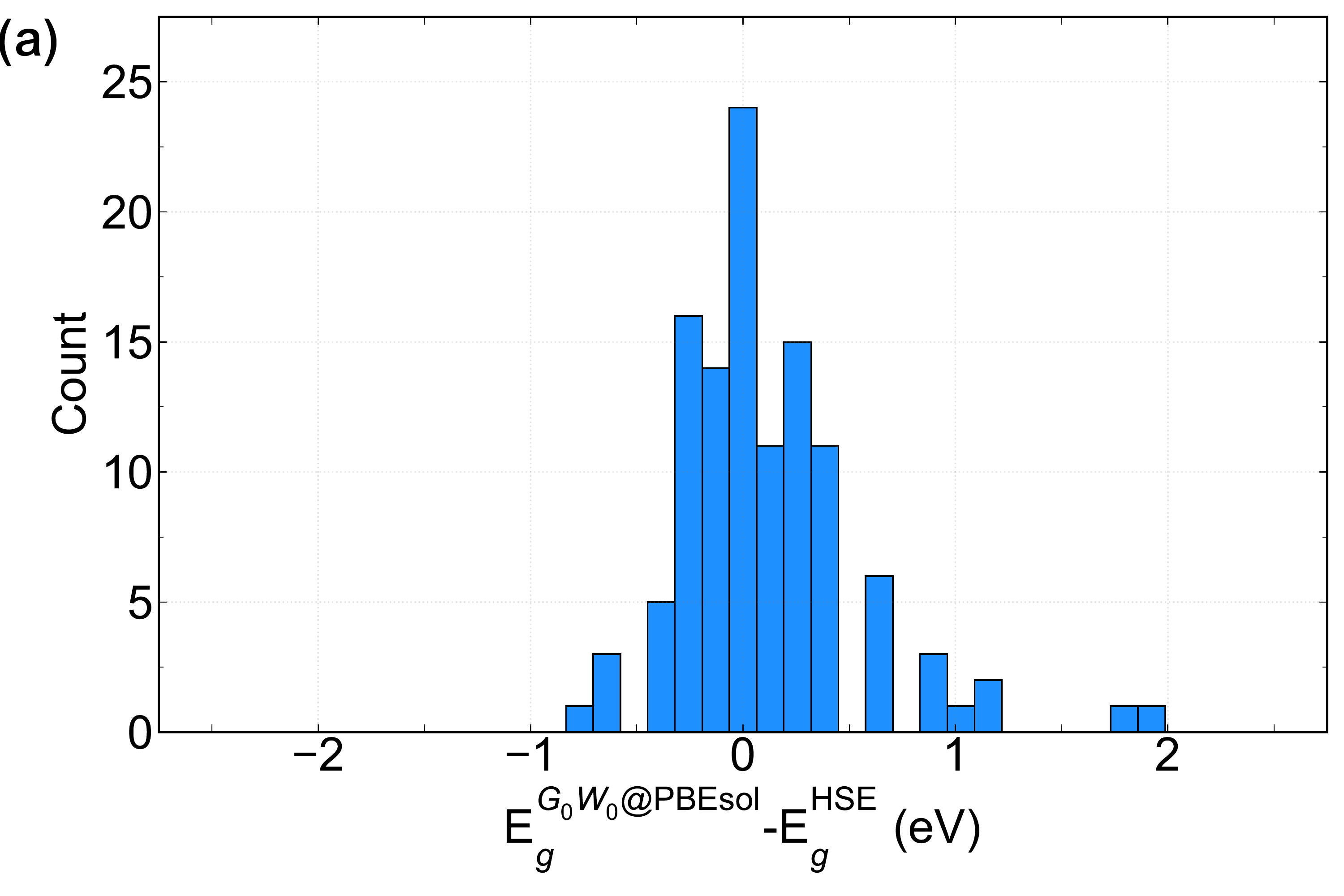}
    \includegraphics[scale=0.25]{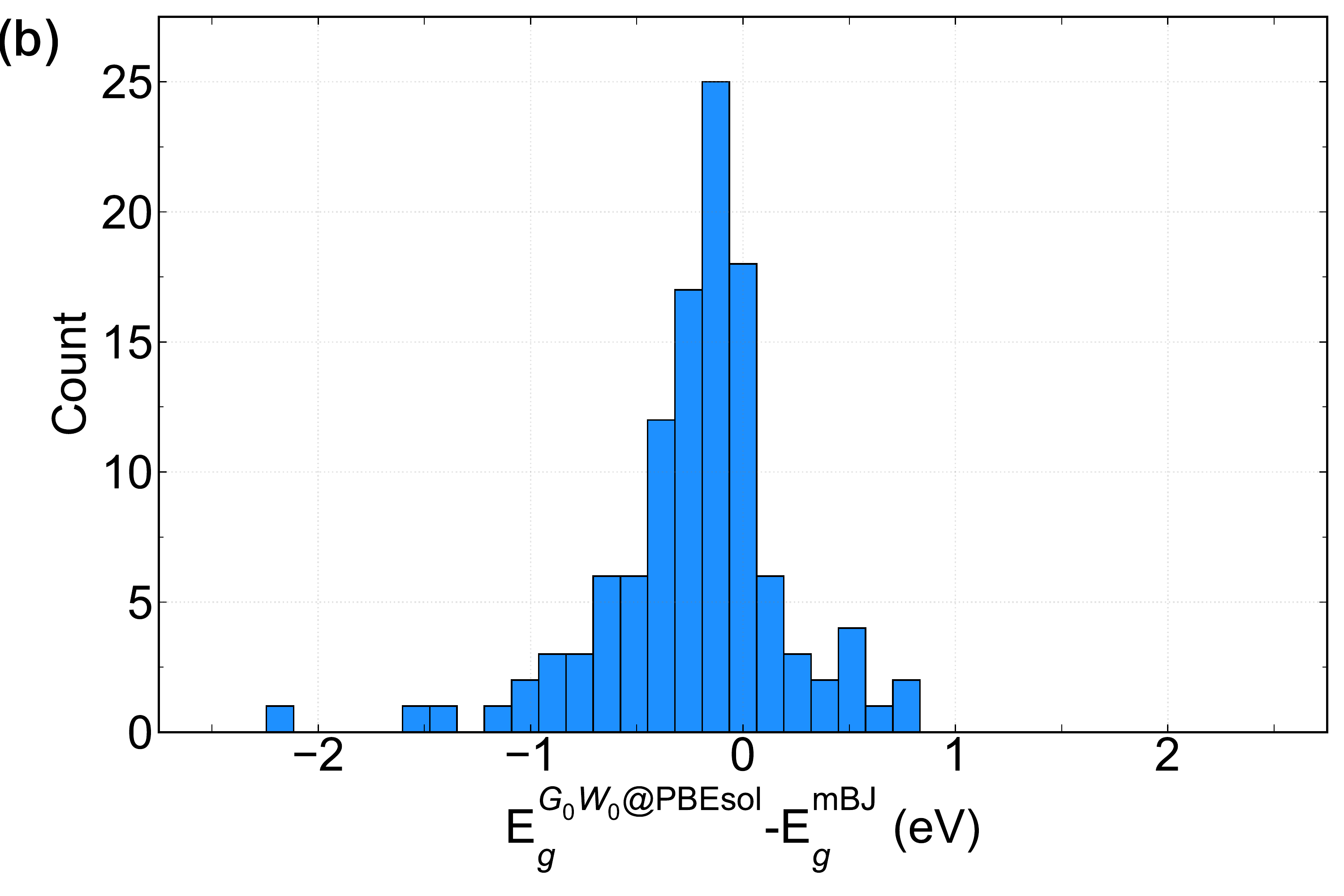} \\
    \includegraphics[scale=0.25]{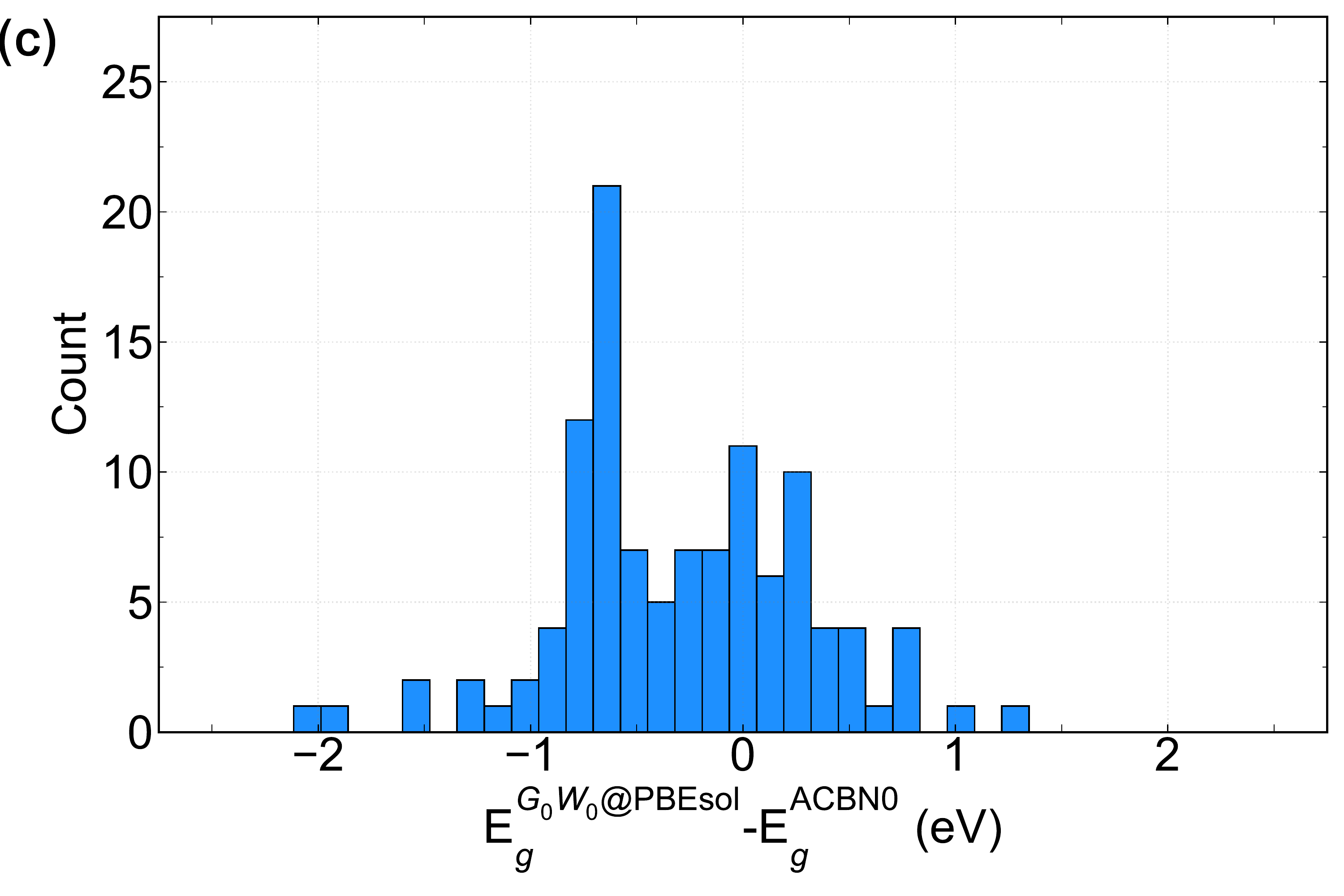}
    \includegraphics[scale=0.25]{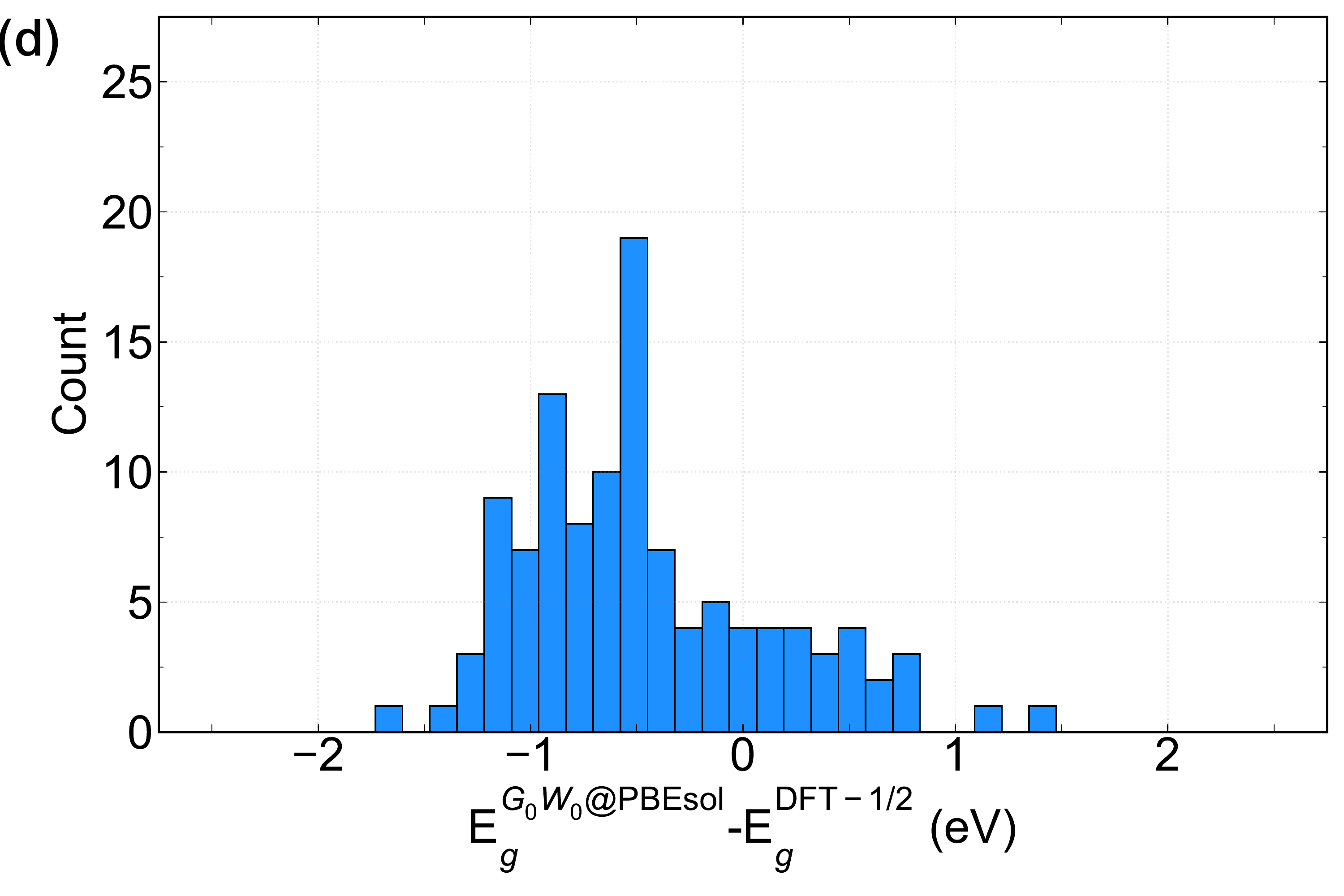} \\
    \caption{
             Distribution histogram of differences between \gw\ and other theoretical methods: (a) HSE06, (b) mBJ, (c) ACBN0, and (d) DFT$-1/2$. 
            } 
\end{figure*}

\begin{figure*}[!ht]
    \centering
    \includegraphics[scale=0.25]{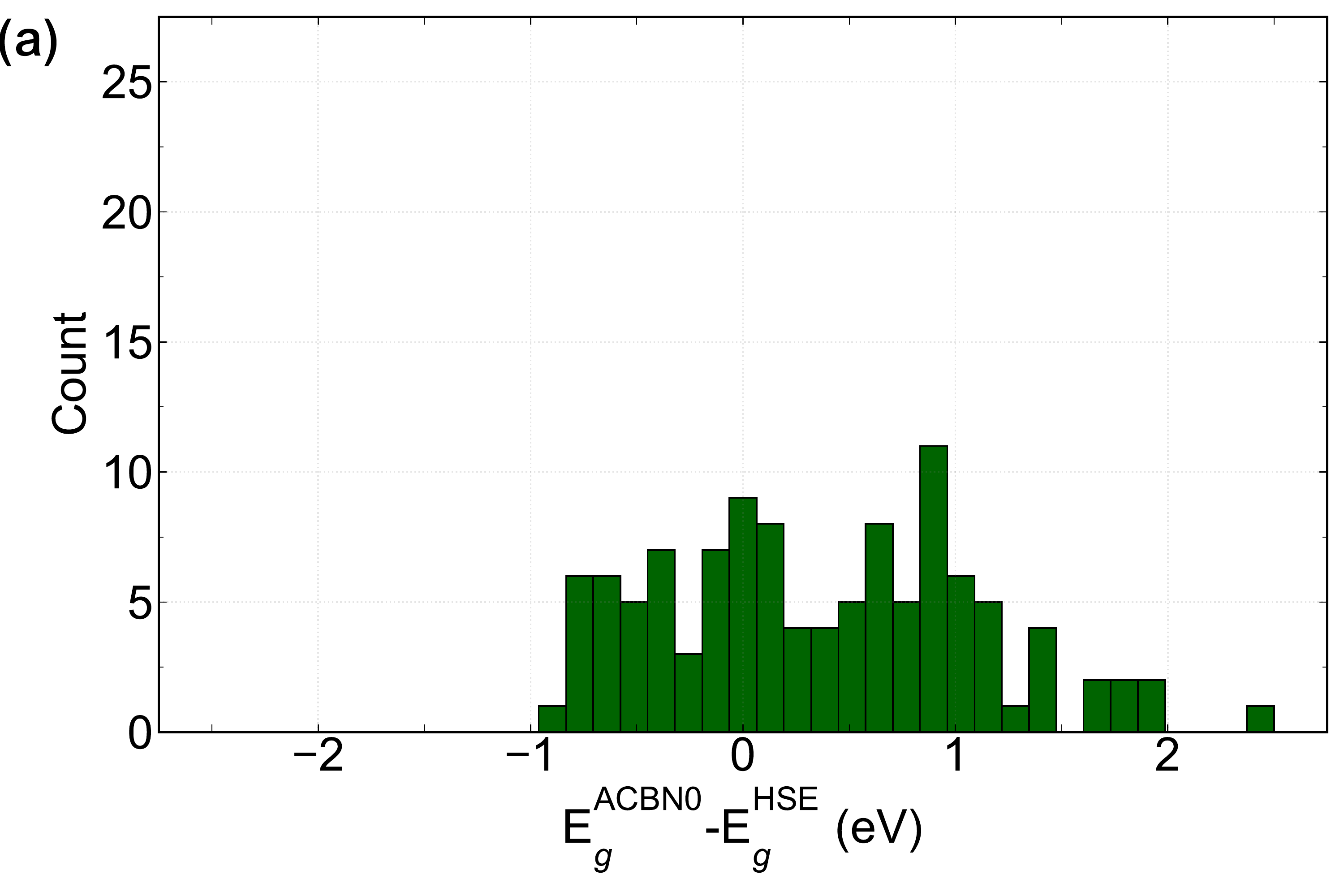}
    \includegraphics[scale=0.25]{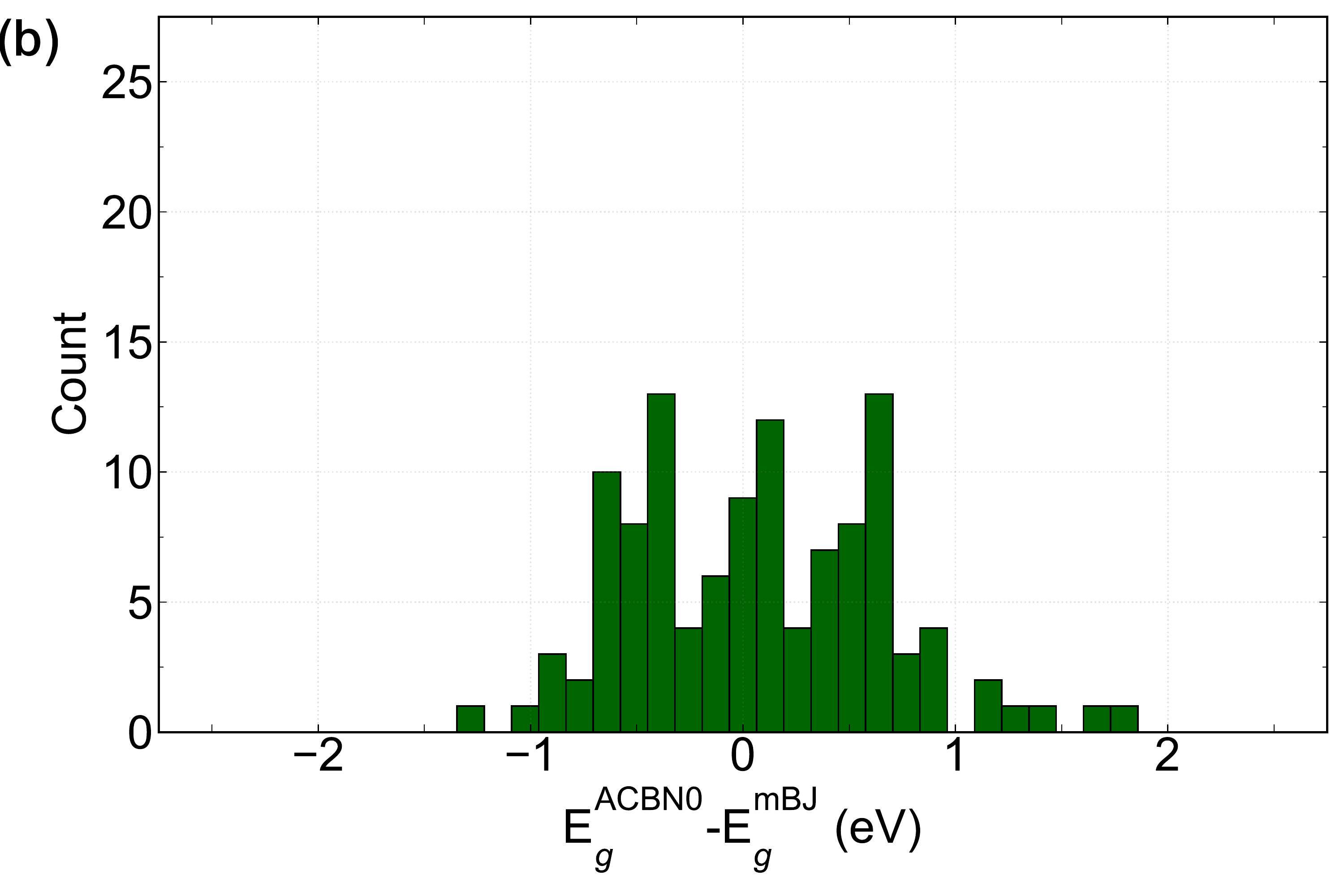} \\
    \includegraphics[scale=0.25]{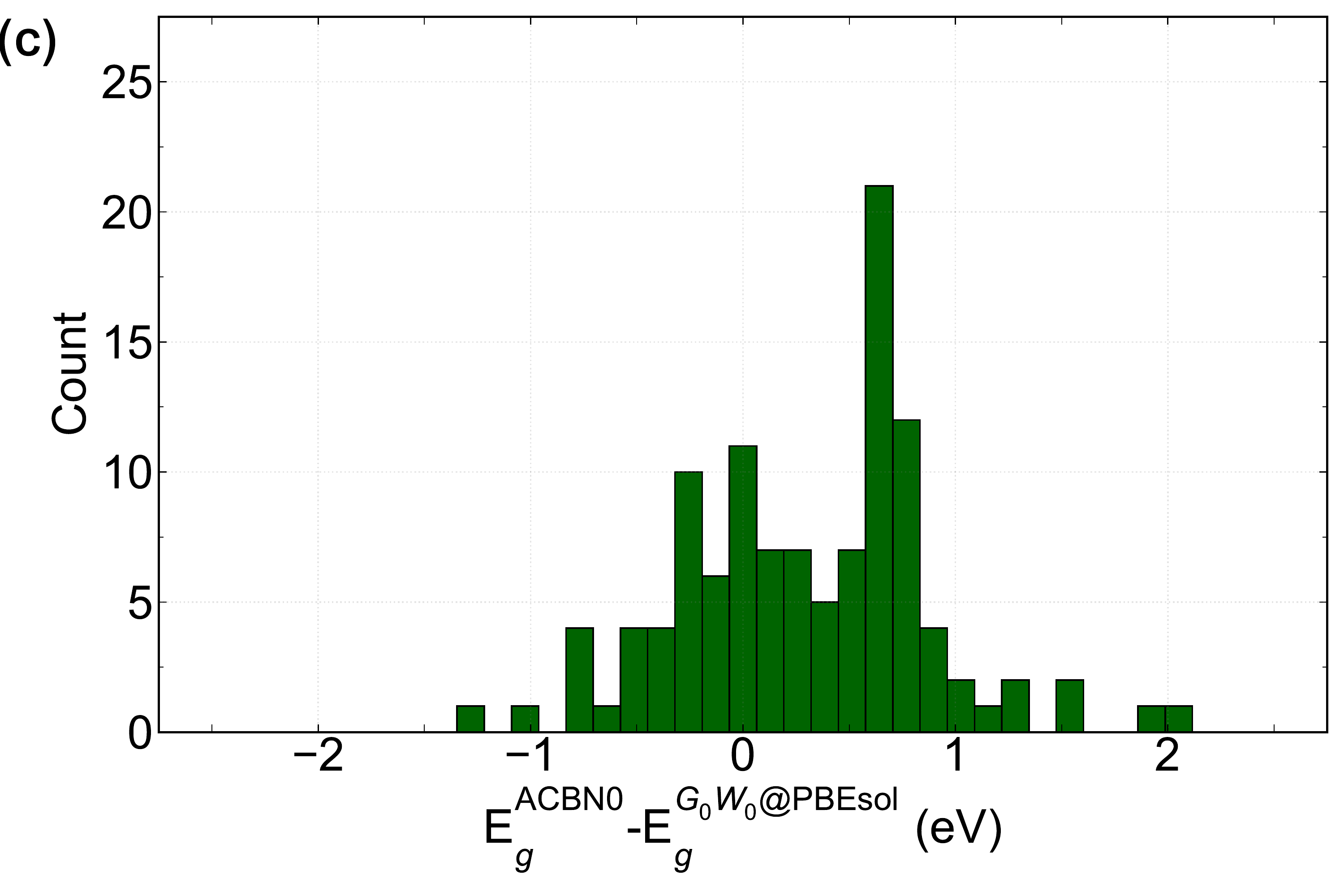}
    \includegraphics[scale=0.25]{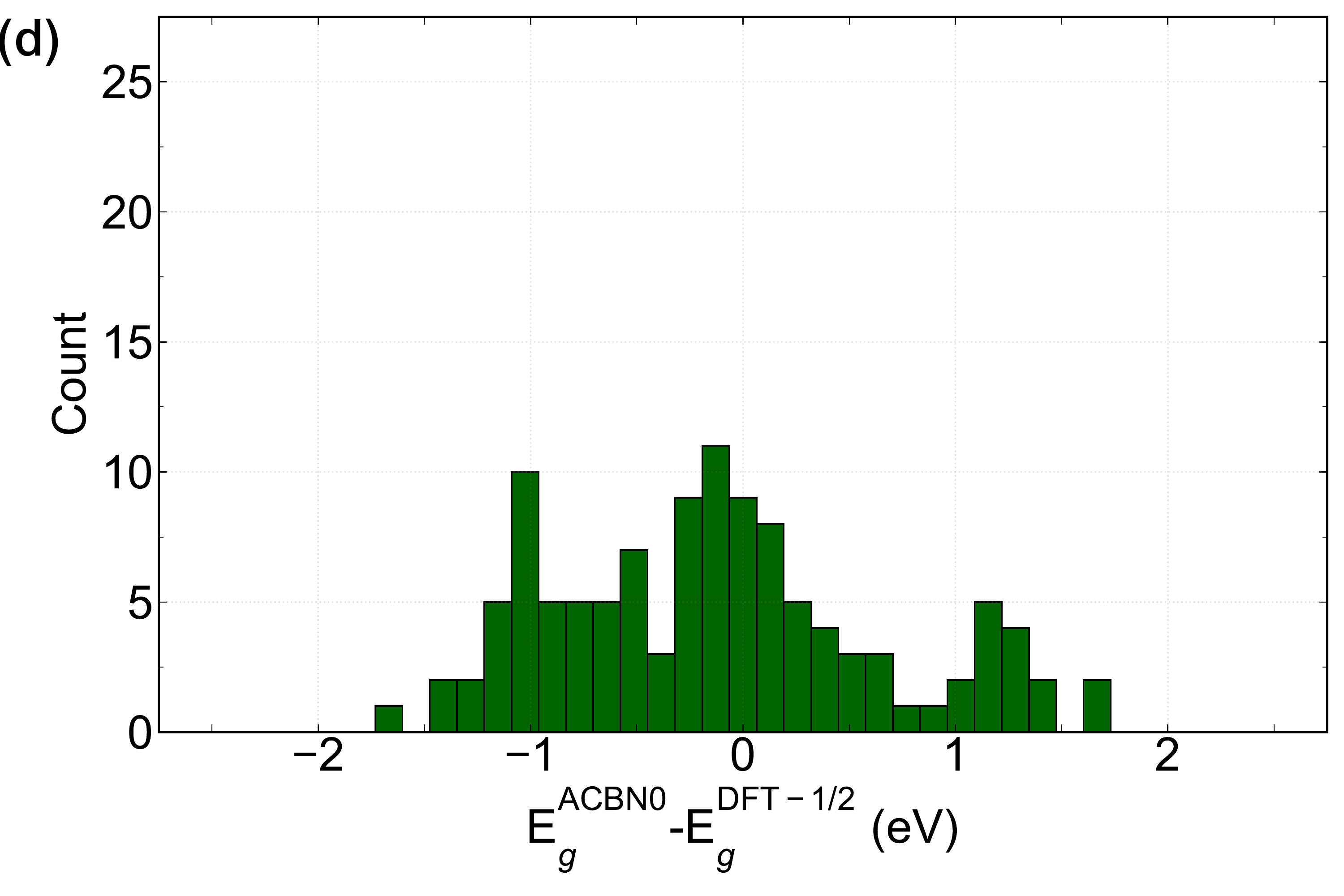} \\
    \caption{
             Distribution histogram of differences between ACBN0 and other theoretical methods: (a) HSE06, (b) mBJ, (c) $G_{0}W_{0}@PBEsol$, and (d) DFT$-1/2$. 
            } 
\end{figure*}

\begin{figure*}
    \centering
    \includegraphics[scale=0.45]{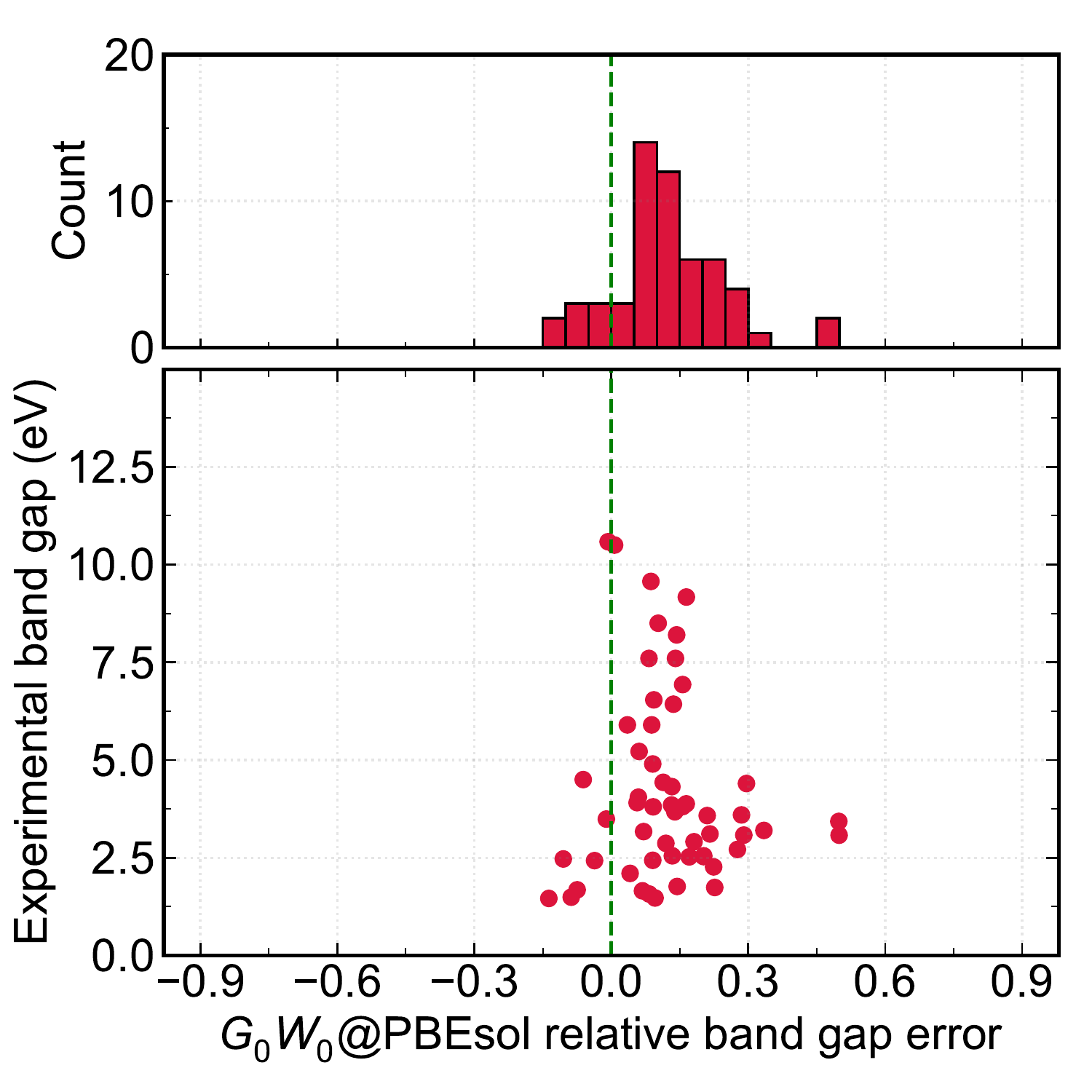}
    \includegraphics[scale=0.45]{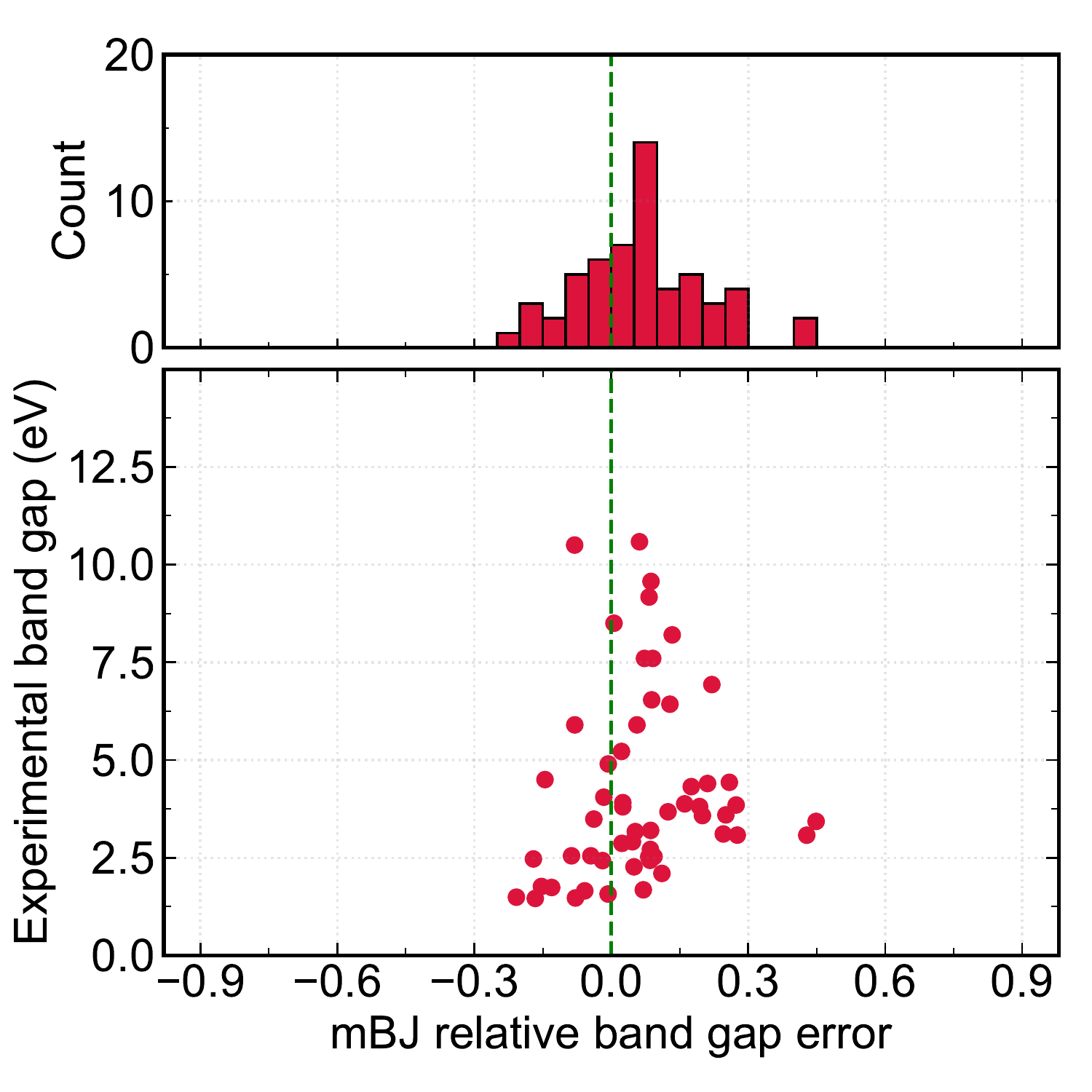} \\
    \includegraphics[scale=0.45]{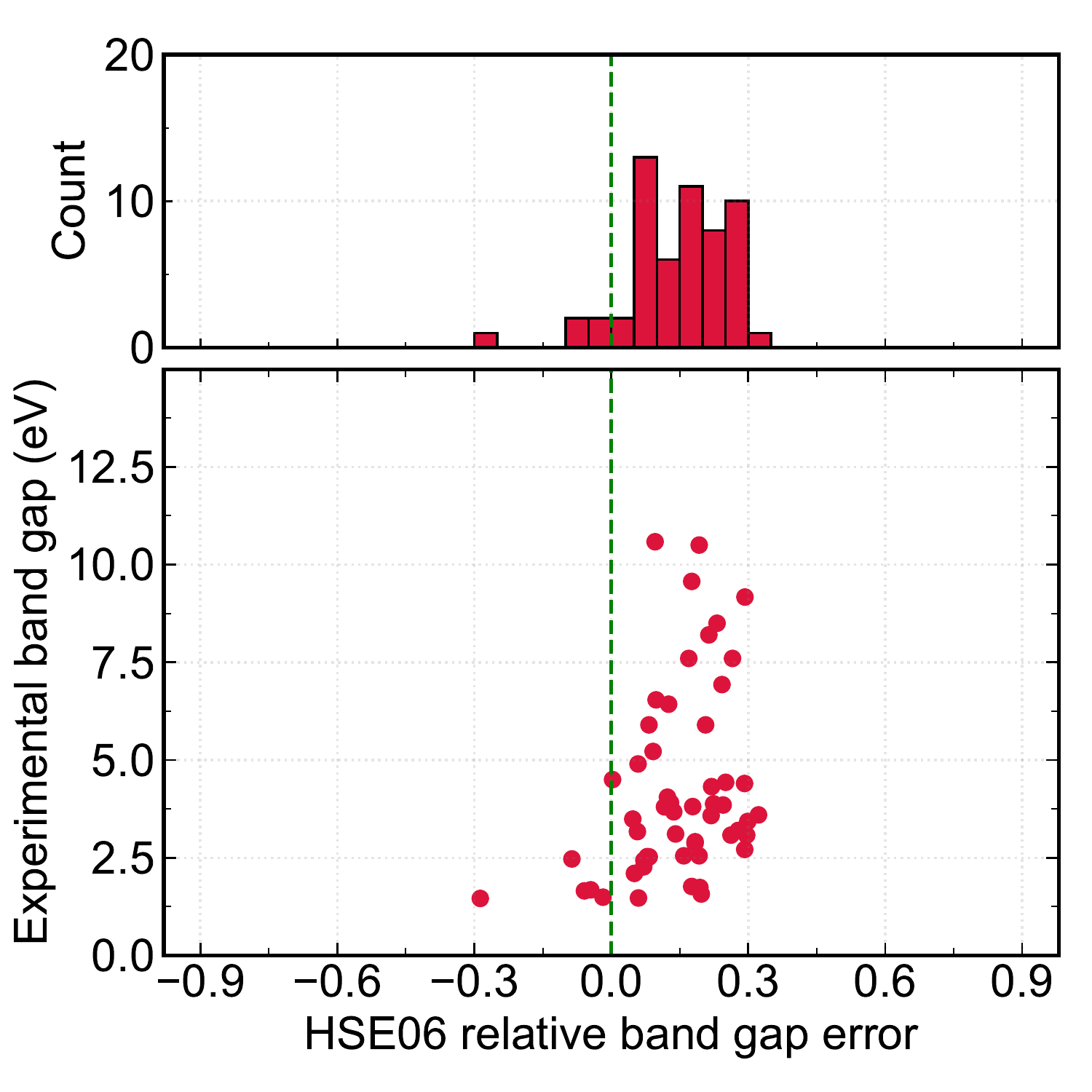}
    \includegraphics[scale=0.45]{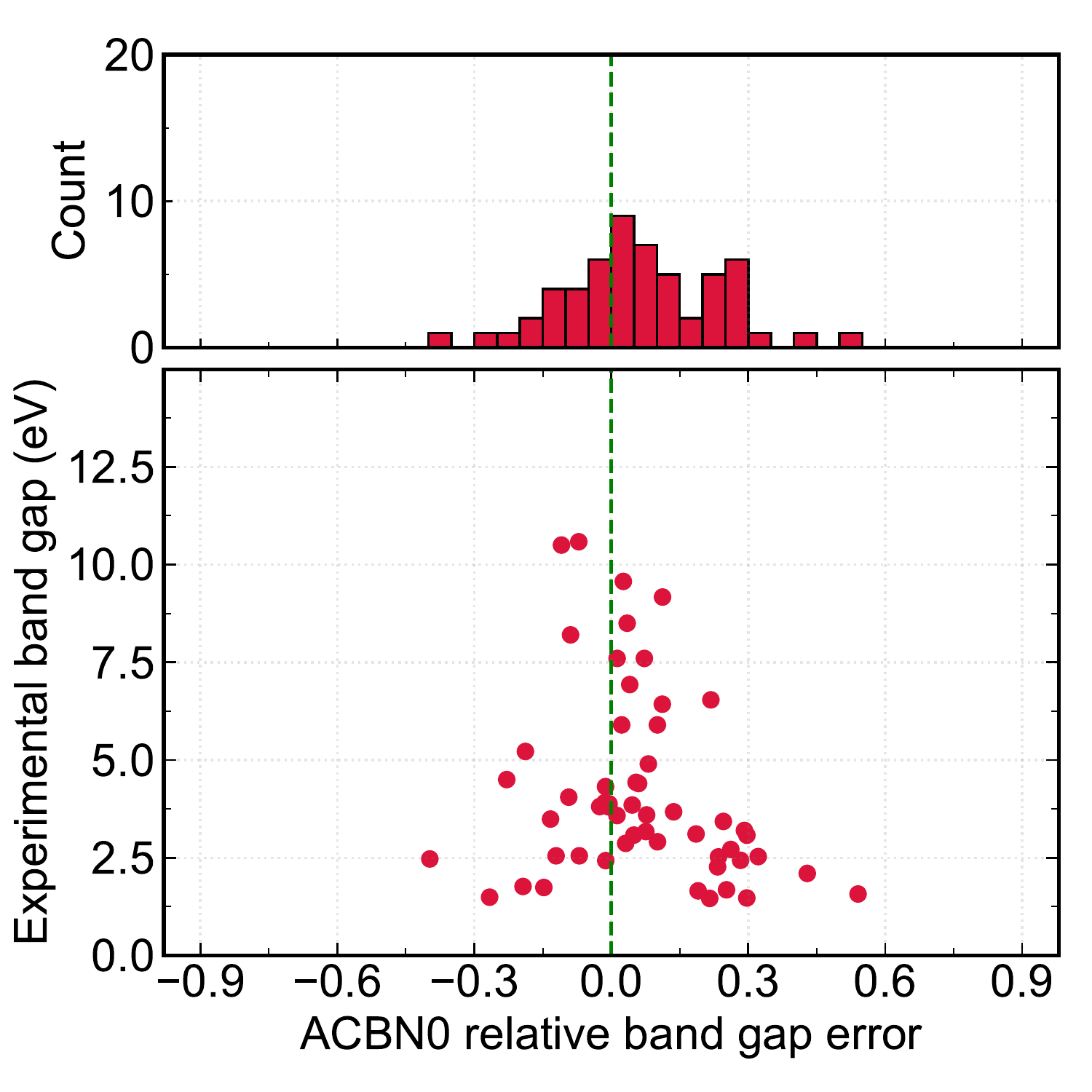} \\
    \includegraphics[scale=0.45]{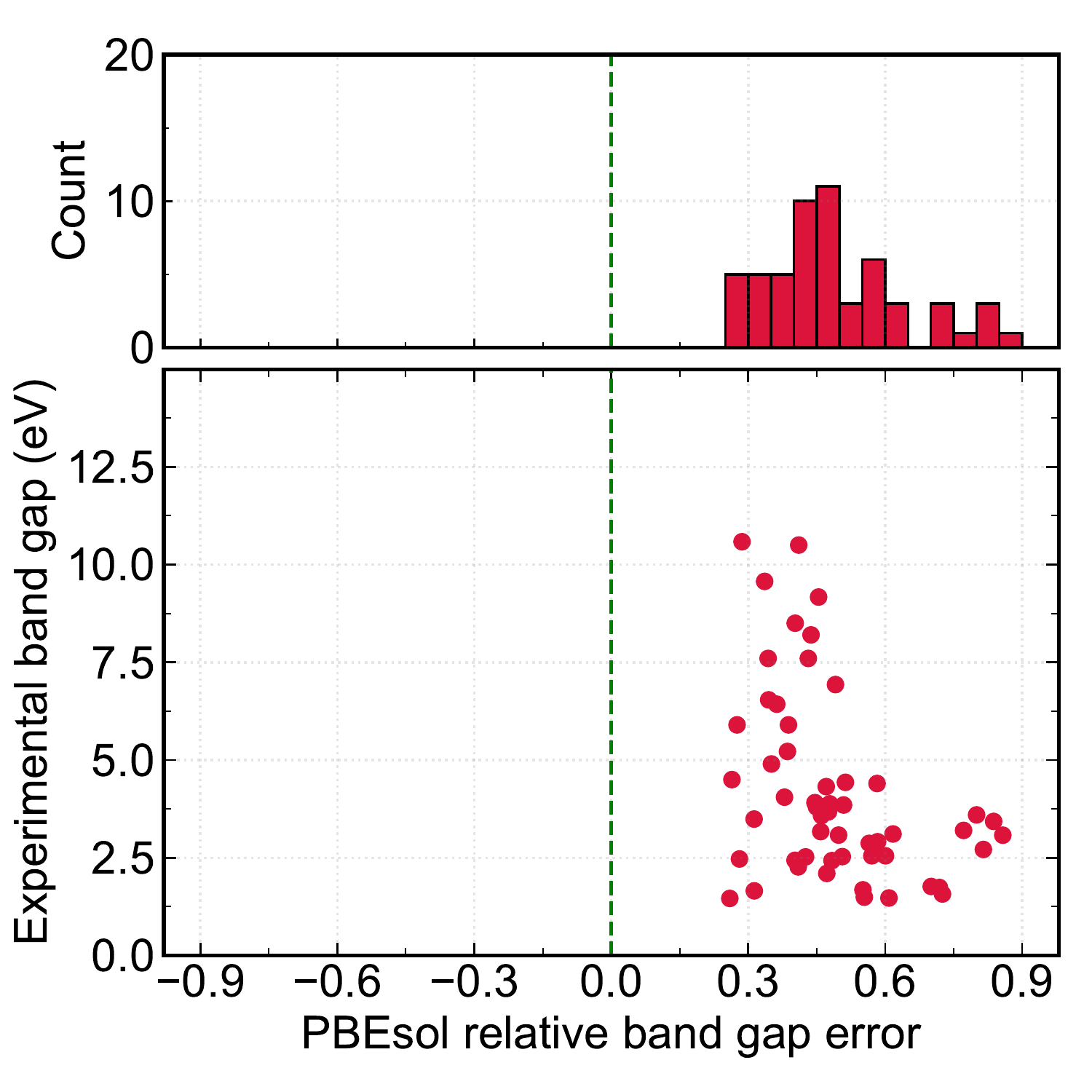}
    \includegraphics[scale=0.45]{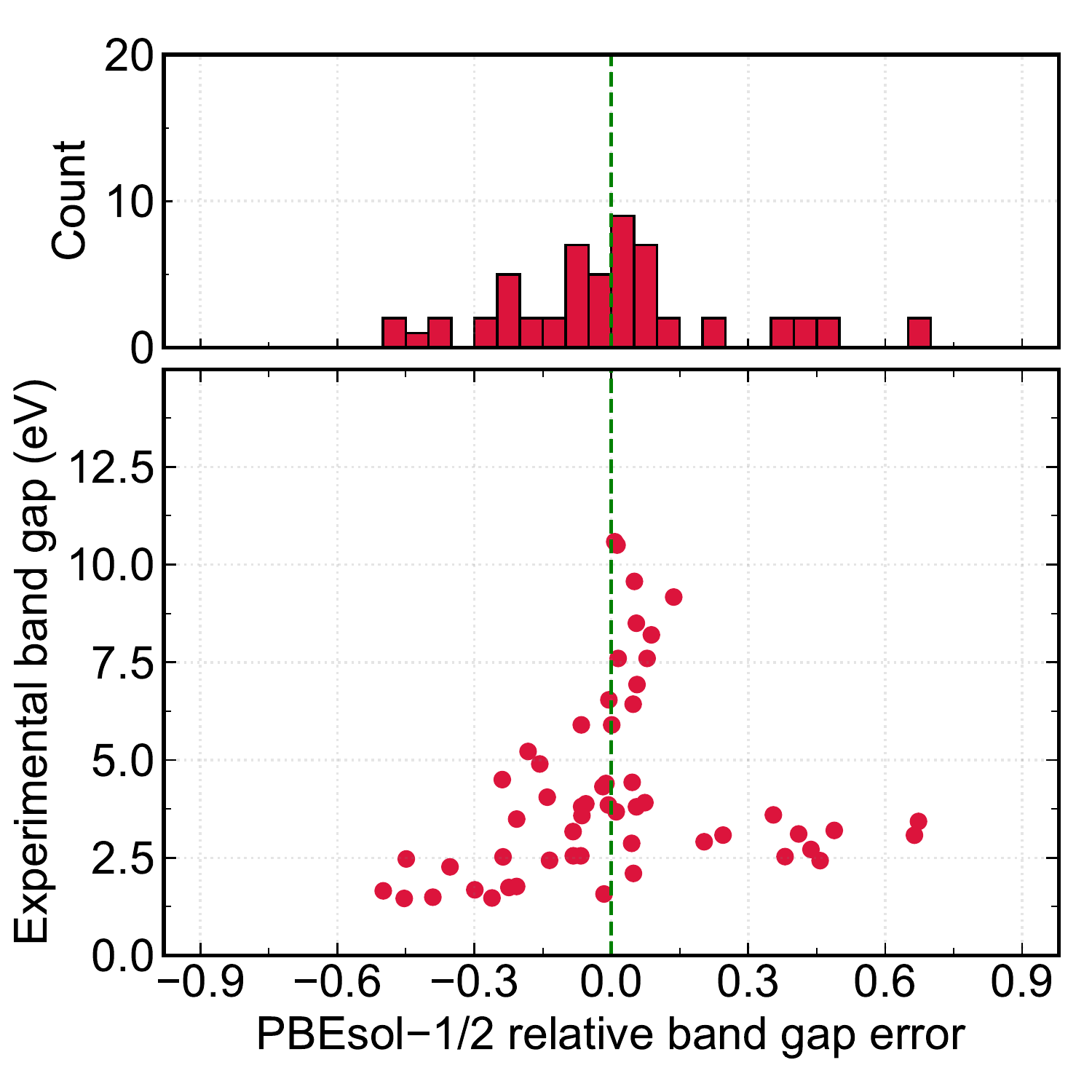} \\
    \caption{
             The relative error between experimental and theoretical band gaps. Experimental band gap plotted as a function of the relative error for (a) $G_{0}W_{0}$@PBEsol, (b) mBJ, (c) HSE06, (d) ACBN0, (e) PBEsol, and (f) PBEsol$-1/2$. The distribution histogram of the relative errors is also presented for each method. The dashed vertical green line corresponds to a zero error value.
            } 
\end{figure*}

\begin{figure*}[]
    \centering
    \includegraphics[scale=0.62]{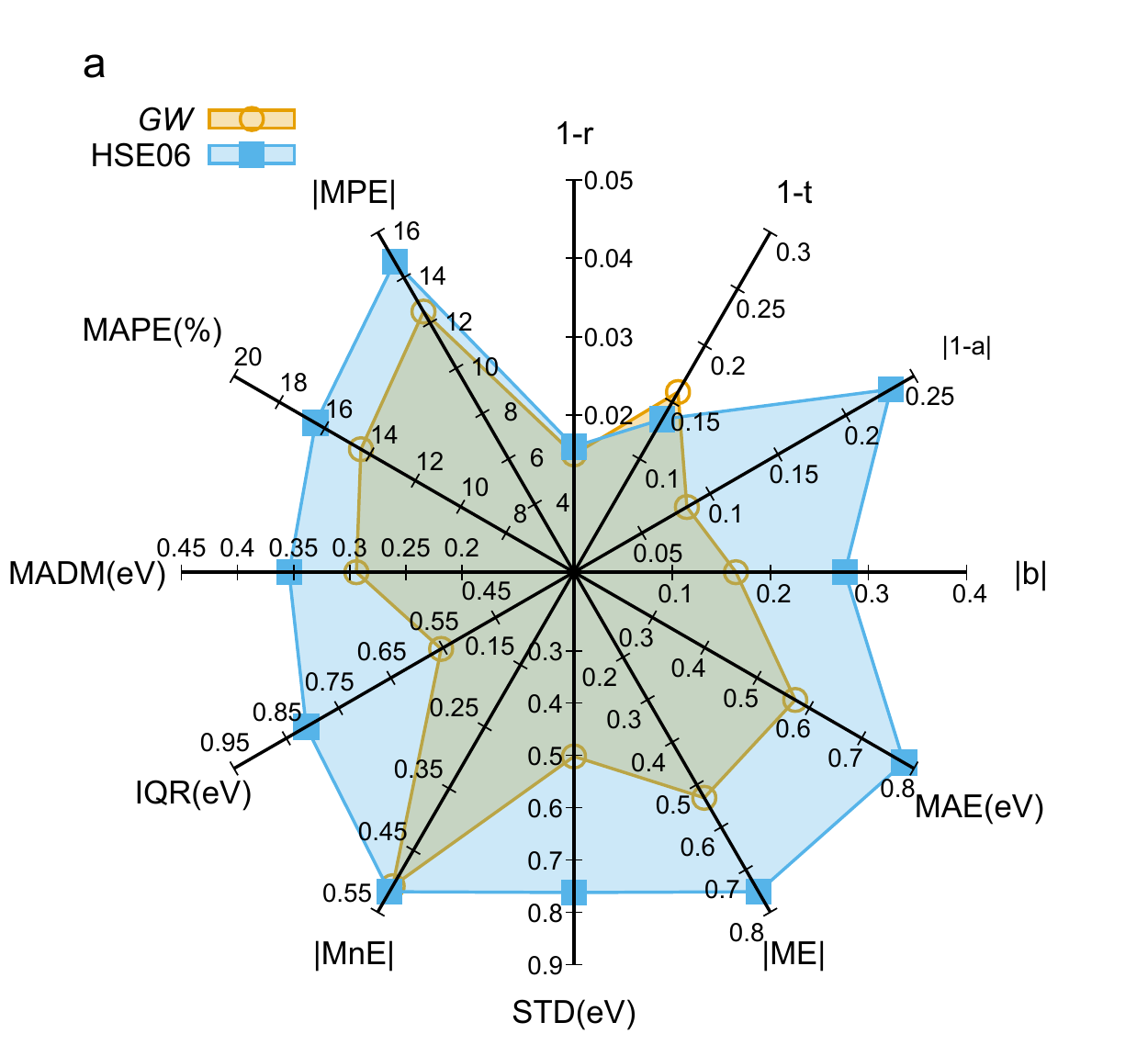}
    \includegraphics[scale=0.62]{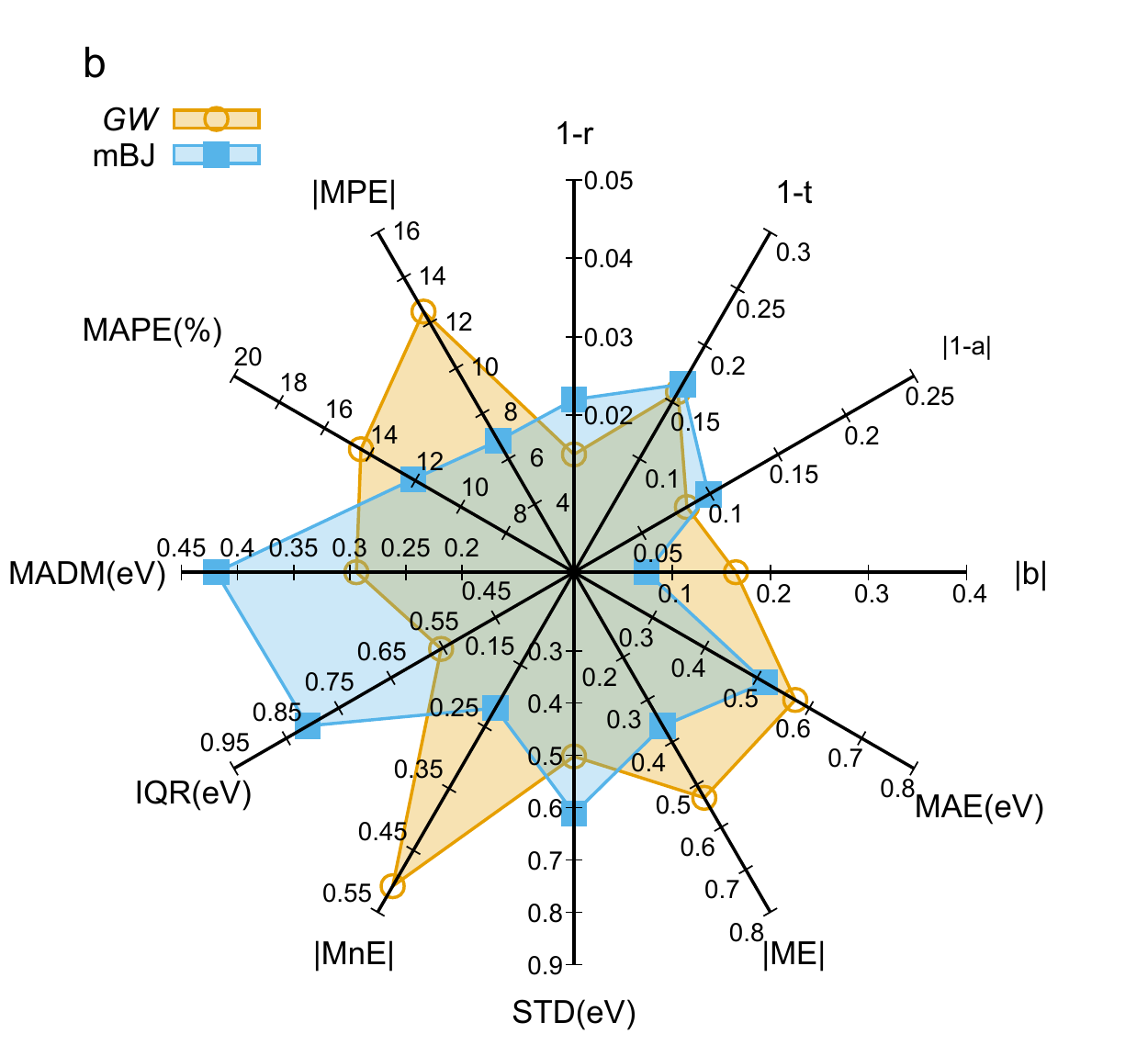} \\
    \includegraphics[scale=0.62]{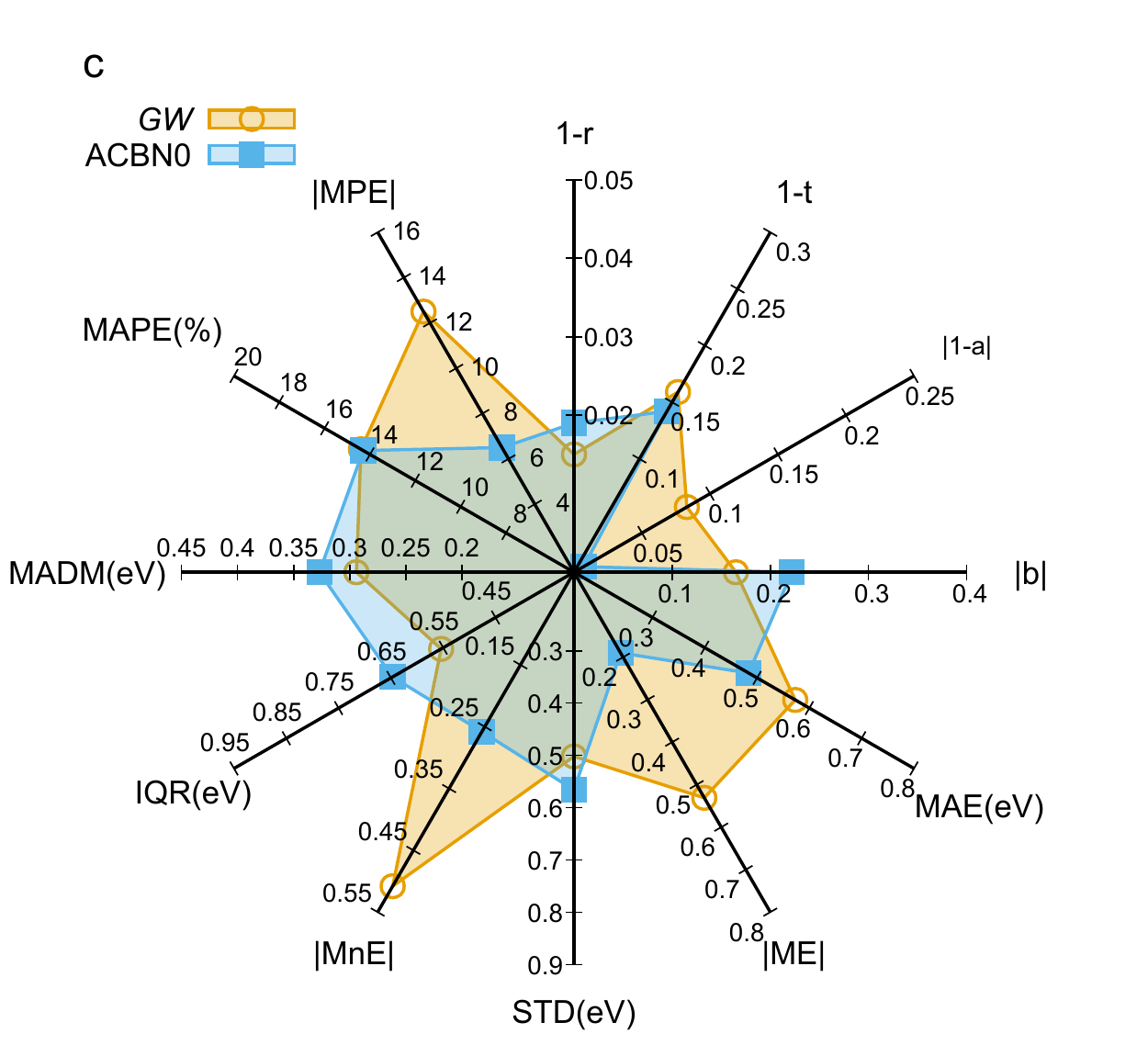} 
    \includegraphics[scale=0.62]{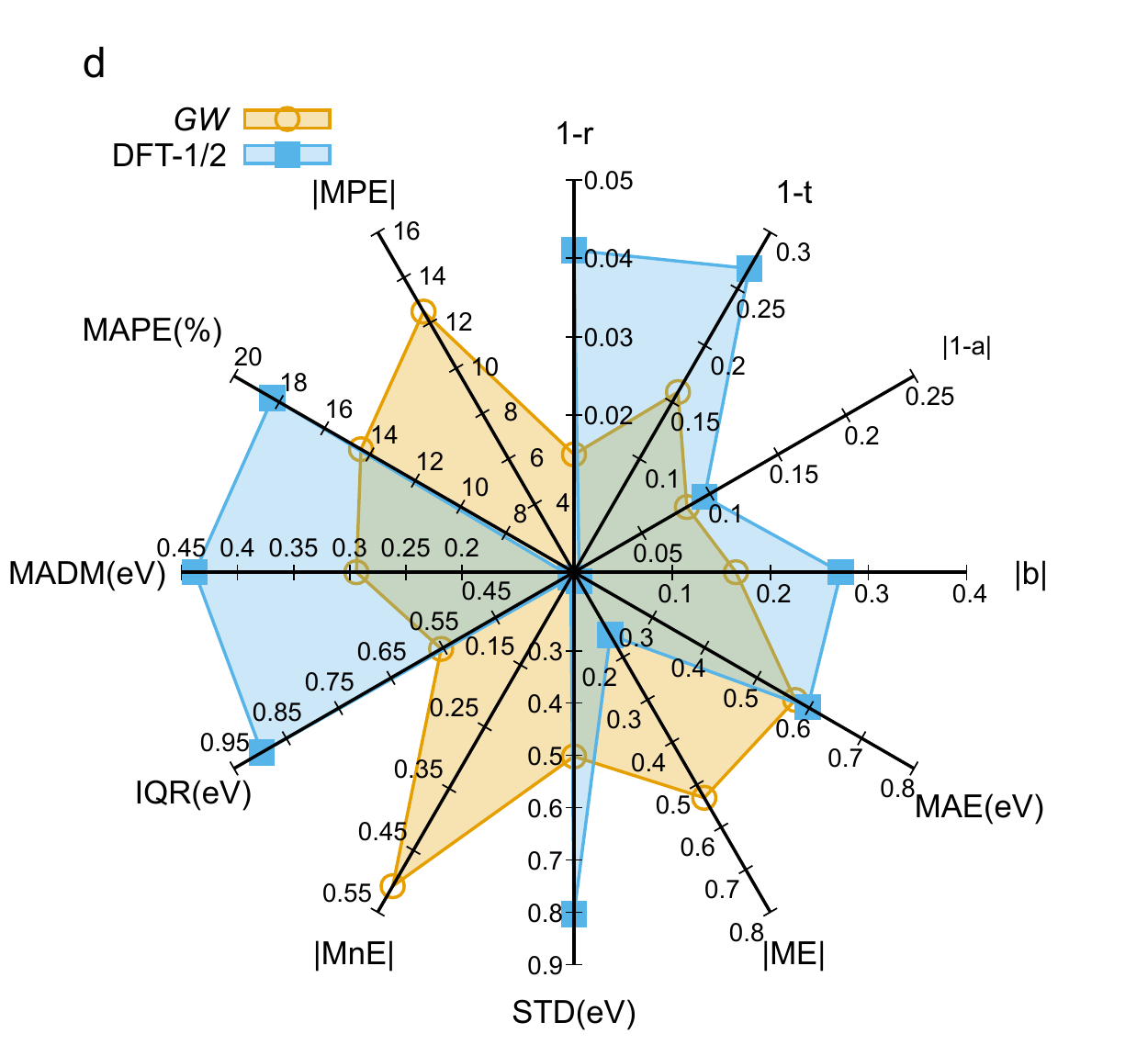}
    \caption{
             Radar plots showing the statistical error quantities calculated in this benchmark study for (a) HSE06, (b) mBJ, (c) ACBN0, and (d) PBEsol$-1/2$. The values of the measures are summarized in Table I in the main text. For a better comparison, the radar plot corresponding to $G_{0}W_{0}$@PBEsol error measures is shown in orange color in each plot.
            } 
\end{figure*}

\begin{figure*}[]
    \centering
    \includegraphics[scale=0.62]{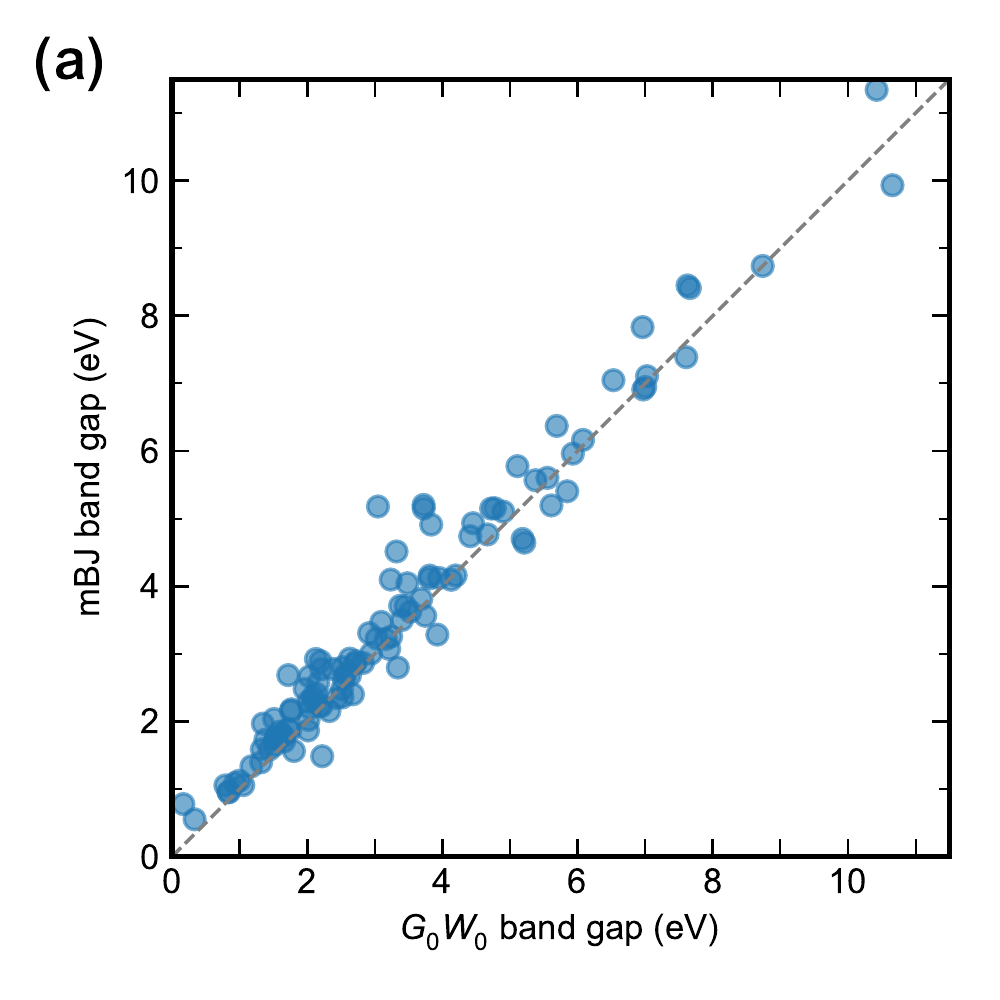}
    \includegraphics[scale=0.62]{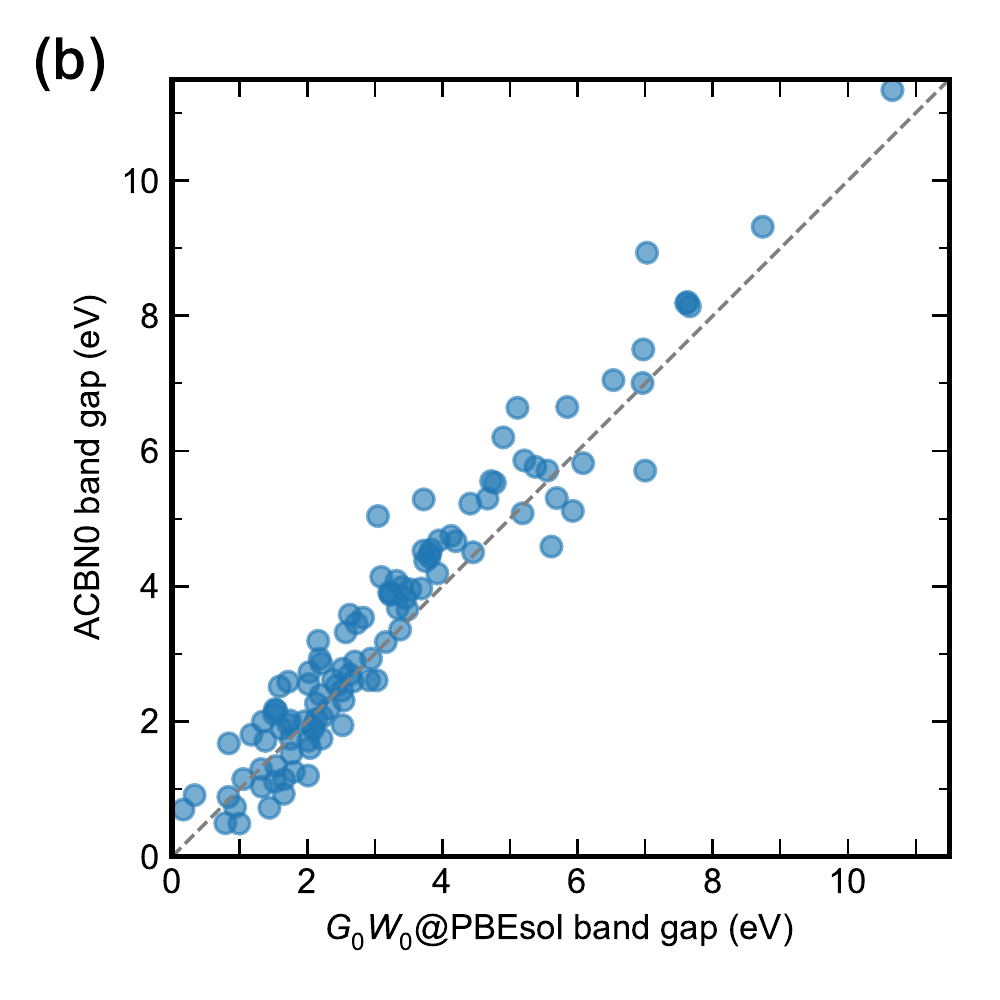} \\
    \includegraphics[scale=0.62]{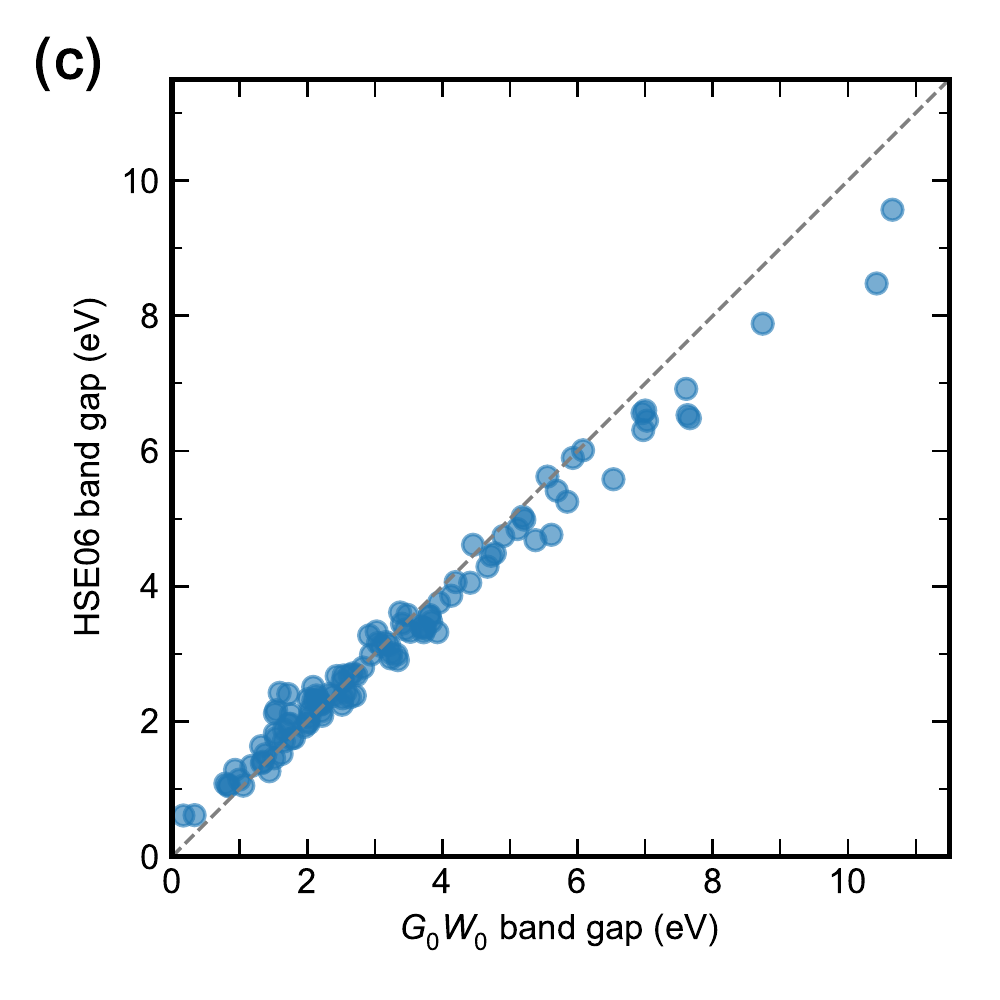}
    \includegraphics[scale=0.62]{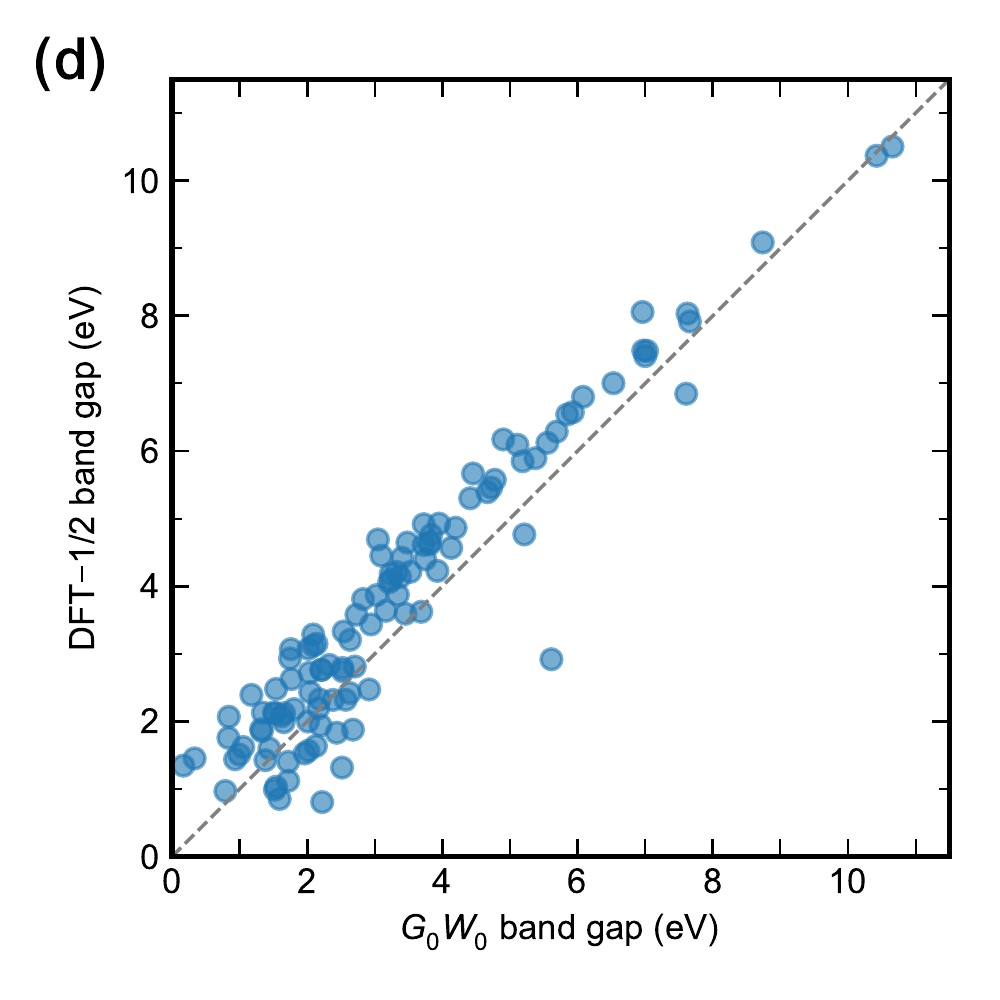}
    \caption{
             Relationships between the quasi-particle band gap from the $G_{0}W_{0}$@PBEsol calculation with (a) HSE06, (b) mBJ, (c) ACBN0, and PBEsol$-1/2$. 
            } 
\end{figure*}

\begin{figure*}[h]
    \centering
    \includegraphics[scale=0.48]{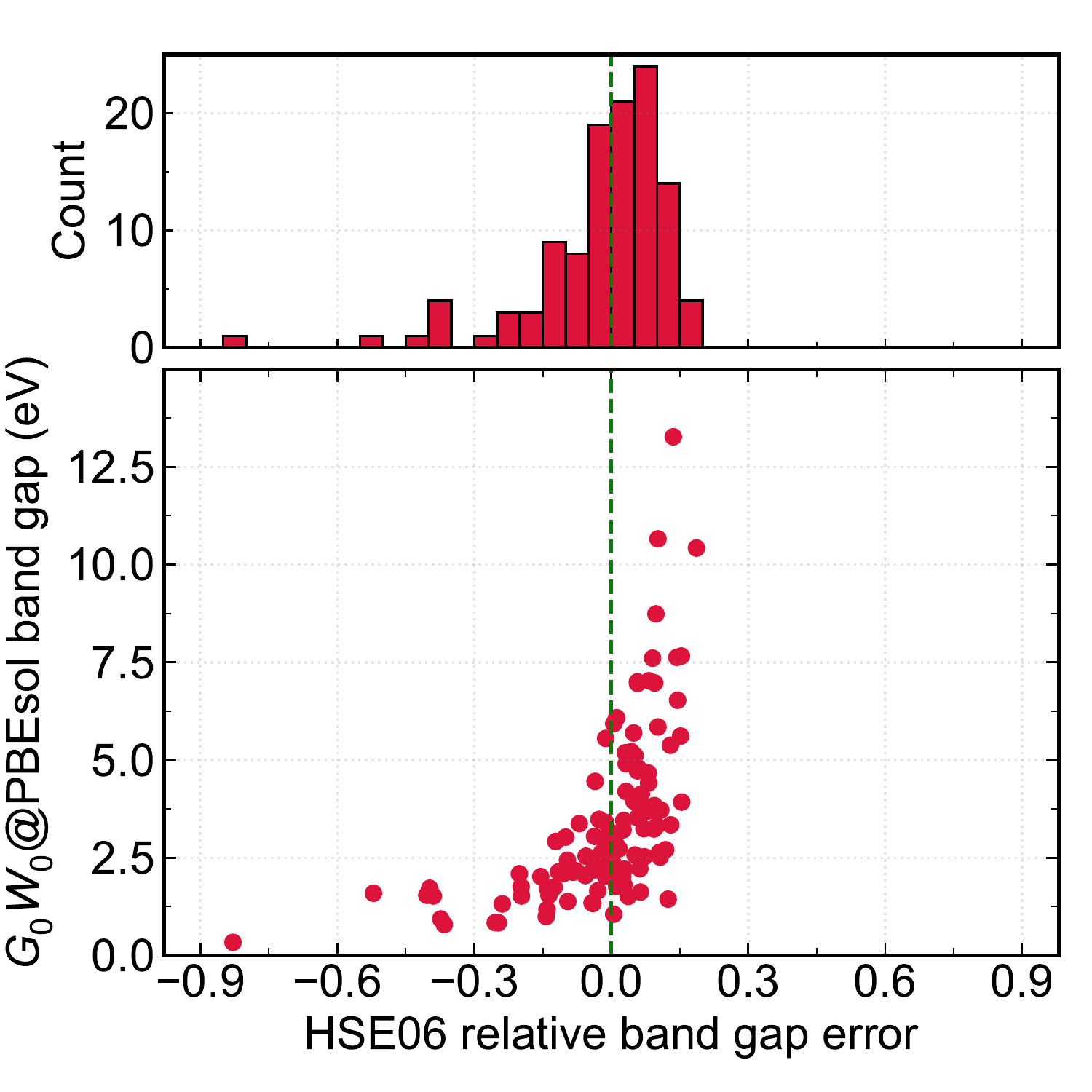}
    \includegraphics[scale=0.48]{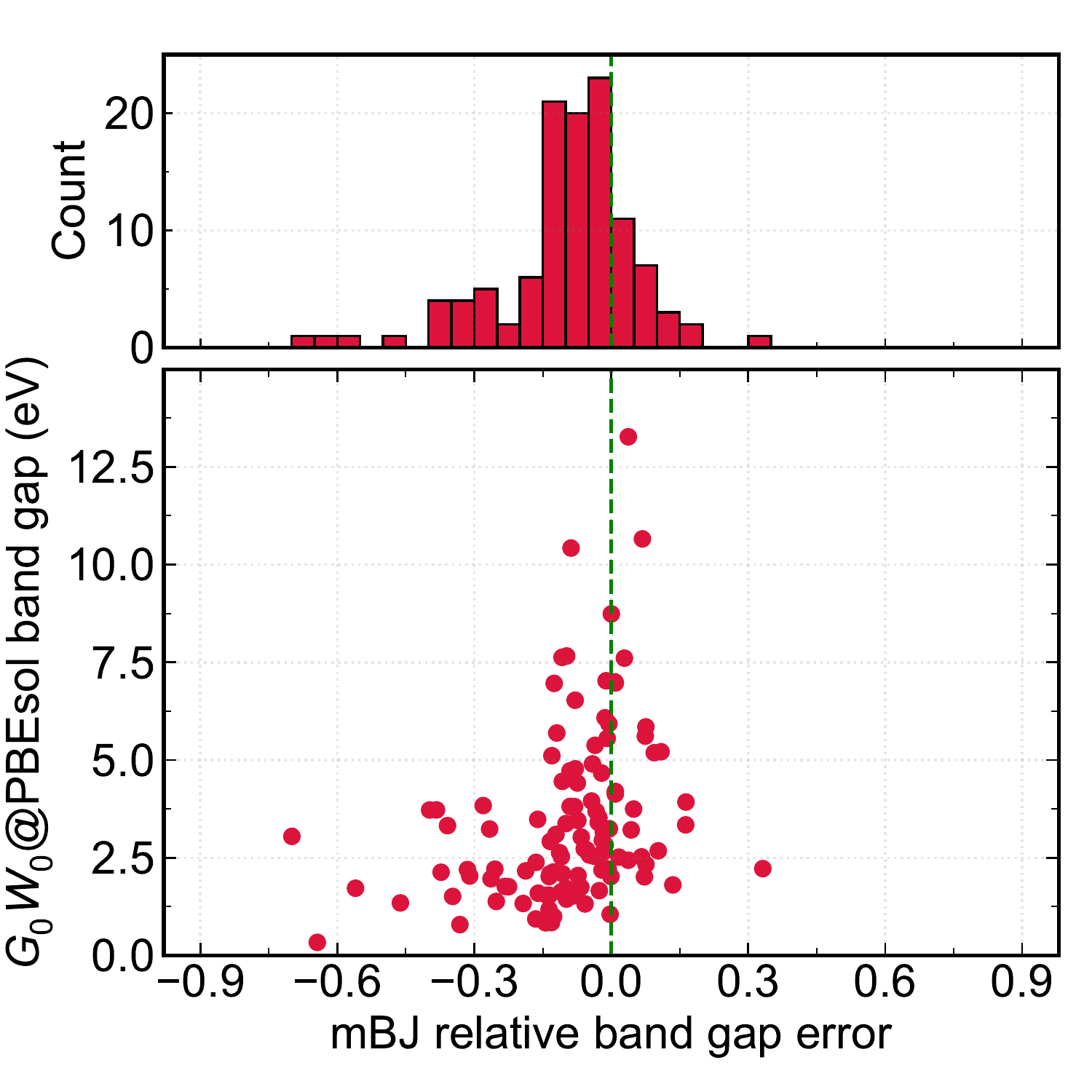} \\
    \includegraphics[scale=0.48]{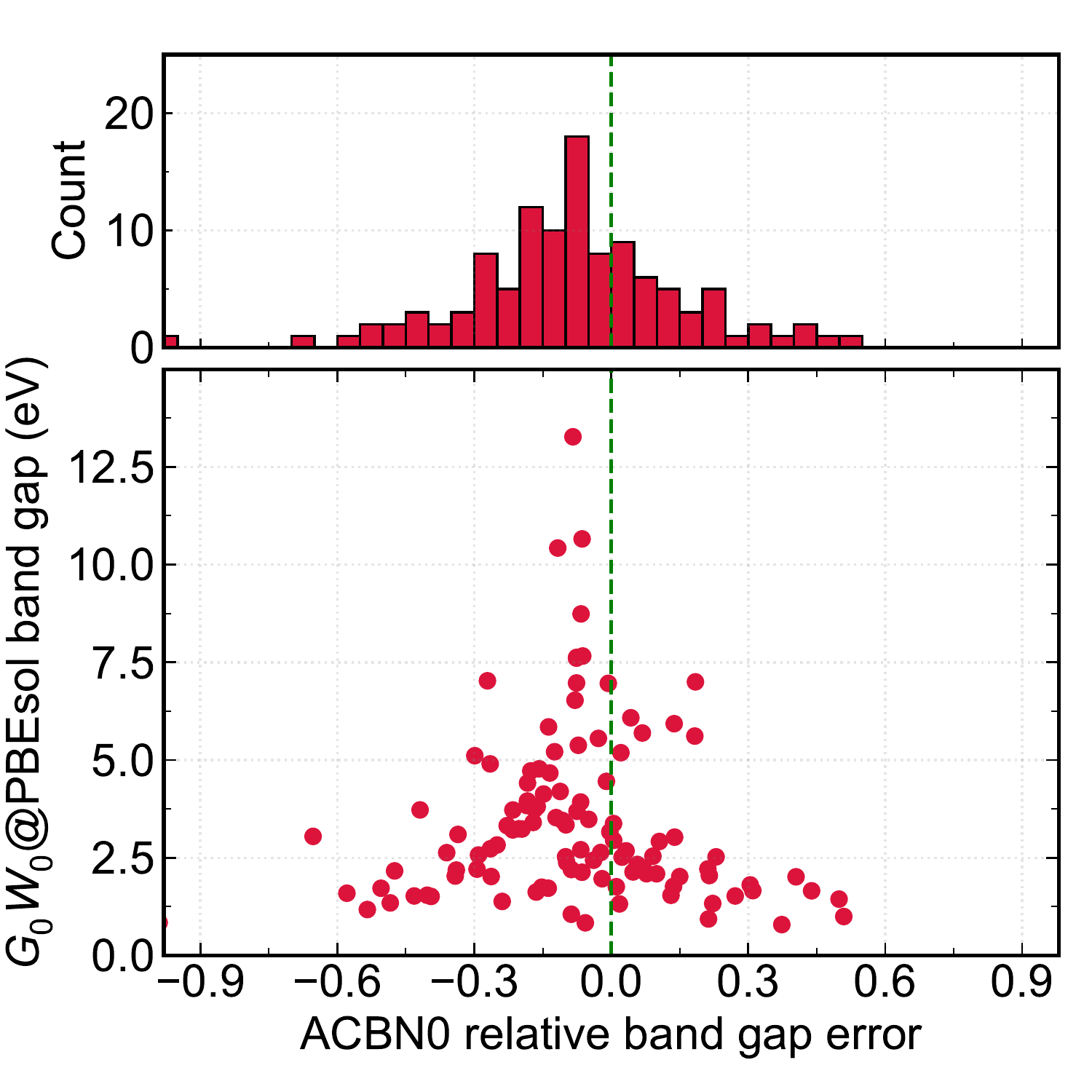}
    \includegraphics[scale=0.48]{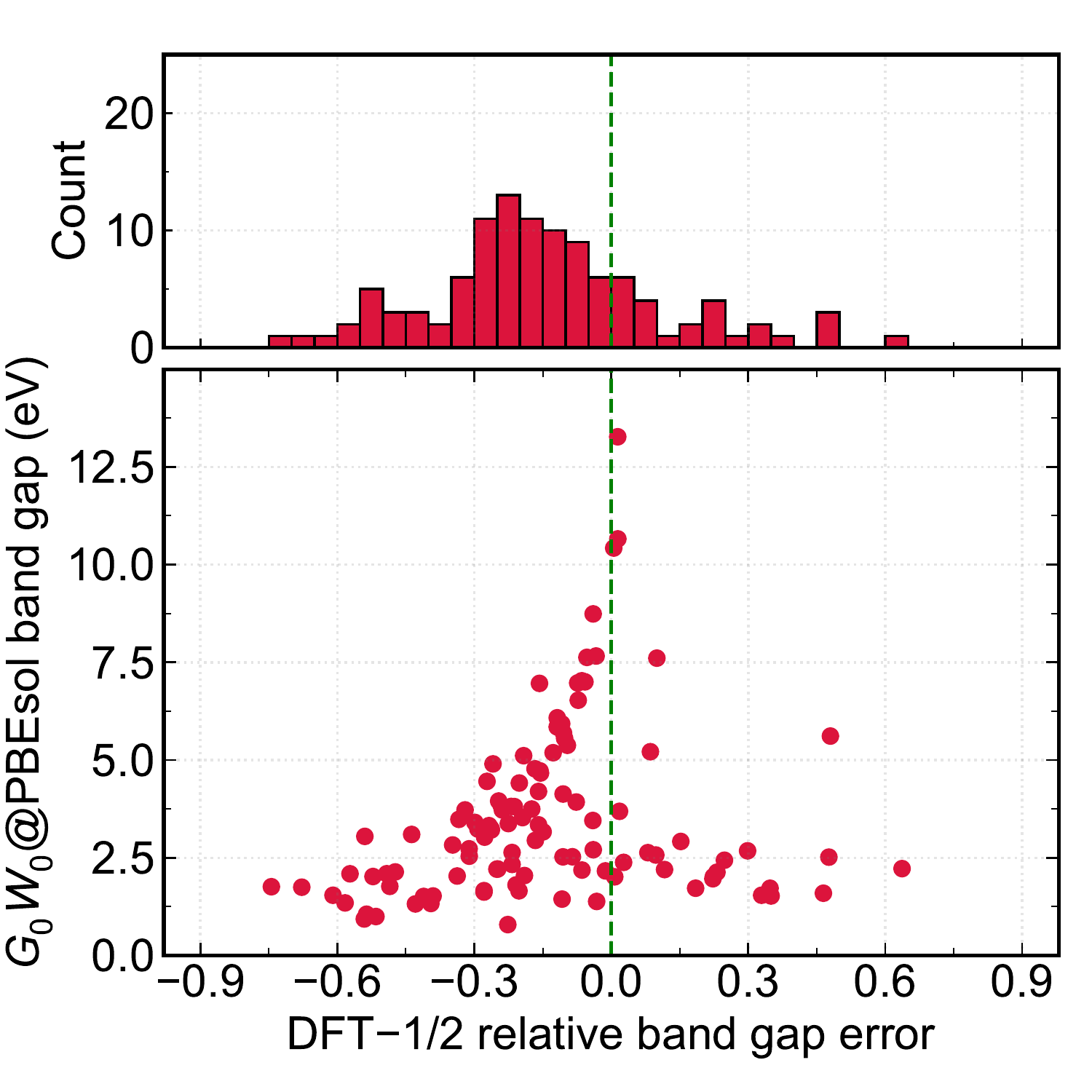} \\
    \caption{
             The relative error between $G_{0}W_{0}$@PBEsol and the band gaps that were obtained from other theoretical methods. $G_{0}W_{0}$@PBEsol band gap plotted as a function of the relative error for (a) HSE06, (b) mBJ, (c) ACBN0, and (d) PBEsol$-1/2$. The distribution histogram of the relative errors is presented above each plot. The dashed vertical green line corresponds to a zero error value.
            } 
\end{figure*}

\begin{table*}[!ht]
\centering
 \caption{
          Statistical error measures of the calculated band gaps within different computational schemes,
          with respect to the experimental data for for a subset after exclusion of the systems with a band gap larger than 7.0\,eV. 
          The blue and black colors show the best and the second-best values, respectively, selected with a slight tolerance (0.01\,eV or 0.1\%).
         }
\begin{ruledtabular}
\newcommand{\bb}{\bf\textcolor{blue}}
\begin{tabular}{lrrrrrr}
{}              & PBEsol &  mBJ       & PBEsol$-1/2$ & ACBN0   & HSE06       & \gw\        \\
\midrule
Pearson r       & 89.4   & 0.928      & 0.868     & 0.936      & \bb{0.959}  & {\bf0.951}  \\
Kendall $\tau$  & 0.637  & 0.768      & 0.622     & {\bf0.805} & \bb{0.814}  & 0.774       \\
$a$             & 0.68   & 0.85       & \bb{0.98} & \bb{0.98}  & 0.81        & {\bf0.89}   \\
$b$ (eV)        & -0.56  & 0.22       & \bb{0.04} & -0.16      & 0.15        & {\bf-0.08}  \\
ME (eV)         & -1.65  & -0.3       & -0.05     & -0.22      & -0.51       & -0.46       \\
MPE (\%)        & -50.6  & -6.8       & 0.6       & -7.6       & -13.4       & -13.1       \\
STD (eV)        & 0.66   & 0.53       & 0.79      & 0.52       & \bb{0.43}   & {\bf0.44}   \\
MAE (eV)        & 1.65   & \bb{0.44}  & 0.58      & \bb{0.45}  & 0.54        & \bf0.51     \\
MAPE (\%)       & 50.6   & \bb{12.8}  & 20.6      & 15.8       & 15.6        & {\bf15.3}   \\
MnE (eV)        & -1.62  & -0.17      & 0.08      & -0.26      & -0.49       & -0.44       \\
IQR (eV)        & 0.87   & 0.81       & 0.74      & \bf0.64    & {\bf0.63}   & \bb{0.41}   \\
MADM (eV)       & 0.45   & {\bf0.30}  & 0.39      & 0.32       & {\bf0.31}   & \bb{0.22}   \\
MaxAE           & 3.40   & {\bf1.54}  & 2.31      & \bb{1.43}  & 1.68        & 1.71        \\
\end{tabular}
\end{ruledtabular}
\end{table*}

\clearpage

\end{document}